\documentclass{article}

\usepackage{PRIMEarxiv}

\errorcontextlines\maxdimen

\usepackage[utf8]{inputenc} 
\usepackage[T1]{fontenc}    
\usepackage{url}            
\usepackage{booktabs}       
\usepackage{nicefrac}       
\usepackage{microtype}      
\usepackage{fancyhdr}       
\usepackage{graphicx}
\usepackage{xcolor}
\usepackage{threeparttable}
\usepackage{booktabs}
\usepackage{multirow}
\usepackage{tabularx}
\usepackage{arydshln}
\usepackage{subcaption}
\usepackage{cite}

\usepackage{lineno,hyperref}
\hypersetup{colorlinks, citecolor=blue, linkcolor=red, urlcolor=blue}
\usepackage{amsmath,amsfonts,amssymb}
\usepackage{graphics}
\usepackage{bm}
\usepackage{multirow}
\usepackage{algorithm}
\usepackage{algpseudocode}

\usepackage{mathtools}
\usepackage{siunitx}
\DeclarePairedDelimiterXPP\BigOSI[2]%
  {\mathcal{O}}{(}{)}{}%
  {\SI{#1}{#2}}

\title{Accelerating metamaterial topology optimization using deep super-resolution networks}

\author{Ajendra Singh \\
  Department of Civil Engineering \\
  Indian Institute of Technology Roorkee\\
  \texttt{a\_singh4@ce.iitr.ac.in} \\
  \And
  Shubham Saurabh \\
  Department of Civil Engineering \\
  Indian Institute of Technology Roorkee\\
  \texttt{shubham.ce@sric.iitr.ac.in} \\
  \And
  Abhinav Gupta \\
  Department of Civil and Environmental Engineering\\
  Vanderbilt University, Nashville, USA\\
  \texttt{abhinav.gupta@vanderbilt.edu} \\
  \And
  Rajib Chowdhury \\
  Department of Civil Engineering\\
  Indian Institute of Technology Roorkee\\
  \texttt{rajib.chowdhury@ce.iitr.ac.in} \\
}

\begin{document}

\maketitle

\begin{abstract}
Designing metamaterials for extreme mechanical behavior involves the optimal selection of design parameters. However, identifying these optimal parameters through topology optimization (TO) across a large parametric space requires extensive computational resources. To address this challenge, we propose a novel deep learning framework for metamaterial topology optimization using an enhanced deep super-resolution (EDSR) approach. Generating low-resolution topologies significantly reduces computational cost compared to high-resolution designs. Therefore, an EDSR network is trained to learn the mapping between low- and high-resolution metamaterial topologies. The training dataset is generated using solid isotropic material with penalization (SIMP)-based TO. We demonstrate the proposed approach for the design of mechanical metamaterials targeting objectives such as maximization of bulk modulus, shear modulus, and elastic modulus, and minimization of Poisson’s ratio. Quantitative assessments --including (i) pixel value error, (ii) objective function error, (iii) intersection over union, and (iv) volume fraction error --validate the accuracy of the EDSR-based TO. Our framework predicts high-resolution topologies of size $192 \times 192$ from optimized low-resolution topologies of size $48 \times 48$. Once trained, the proposed network predicts these high-resolution topologies with only $5-7\%$ of the computational cost required by conventional SIMP-based TO at the same resolution. Moreover, by adding upscale blocks, the framework can generate smoother, higher-resolution topologies suitable for 3D printing. This approach offers a scalable and efficient solution with strong potential for multidisciplinary metamaterial design applications.
\end{abstract}


\section{Introduction}
\label{sec:introduction}
Topology optimization (TO) seeks to identify the best material layout for particular boundary conditions by iteratively optimizing the objective function in a design domain. TO has been extensively used since its introduction by \cite{bendsoeGeneratingOptimalTopologies1988a}, and various approaches have been developed, which can be broadly categorized into density-based and heuristic/intuitive approaches. Density-based approaches include methods such as the homogenization method \cite{bendsoeGeneratingOptimalTopologies1988a,bendsoeMaterialInterpolationSchemes1999}, level set-based method \cite{liTopologyOptimizationConcurrent2018,ghasemiMultimaterialLevelSetbased2018,zhangLevelSetbasedTopological2023}, and Solid Isotropic Material with Penalization (SIMP) method \cite{bendsoeMaterialInterpolationSchemes1999,padhi2022efficient,chuTopologyOptimizationMultimaterial2019,areiasCoupledFiniteelementTopology2021, karuthedath2023continuous, gupta2022adaptive}. Heuristic approaches in TO include computer-aided optimization (CAO) \cite{mattheckNewMethodStructural1990}, soft kill method \cite{baumgartnerSKOSoftKill1992}, bidirectional evolutionary structural optimization (BESO) \cite{huangConvergentMeshindependentSolutions2007,huangEvolutionaryTopologyOptimization2010}, and several others \cite{shobeiri2016optimal,mauersberger2023topology}.

TO has gained significant popularity recently because of its applications in aerospace \cite{munkBenefitsApplyingTopology2019, zhuTopologyOptimizationAircraft2016a}, fluid-structure interaction problems \cite{yoonTopologyOptimizationStationary2010,andreasenTopologyOptimizationFluid2013}, and multiscale nonlinear structures \cite{xiaRecentAdvancesTopology2017, patelImprovingConnectivityAccelerating2022}. It has also been applied to metamaterial microstructure designs \cite{saurabh2023impact,liuTopologicalDesignMicrostructures2021,wiltAcceleratingAuxeticMetamaterial2020a, vangelatosArchitectedMetamaterialsTailored2019b}.

Metamaterials are manufactured structures having extreme properties that do not exist in nature \cite{sigmundSystematicDesignMetamaterials2009}. The anomalous behavior of materials is caused by the periodic arrangement of their microstructures instead of their chemical composition. These materials are presently employed for multidisciplinary applications, including stealth fighters, aerospace, and medical instruments. Recent advancements in metamaterials have led to the design of various forms of mechanical metamaterials, such as those with negative Poisson’s ratio \cite{agrawalRobustTopologyOptimization2022, mizziAuxeticMetamaterialsExhibiting2015a,gattHierarchicalAuxeticMechanical2015b,drosopoulosMechanicalBehaviourAuxetic2016a,zhou2023ready}, negative refractive index \cite{smithMetamaterialsNegativeRefractive2004} and unique thermal expansion properties \cite{wangTopologicalDesignMechanical2016}. Additionally, metamaterials have been developed for vibration control \cite{chenHarnessingMultilayeredSoil2019,drosopoulosEvaluationDynamicResponse2019}, acoustic \cite{chenAcousticCloakingThree2007,noguchiLevelSetbasedTopology2022}, and pentamode metamaterials \cite{liTopologicalDesignPentamode2021}.

For the first time, \cite{sigmundMaterialsPrescribedConstitutive1994} used TO to design the microstructure. Afterward, various TO methods were employed, including the density-based approach \cite{nevesOptimalDesignPeriodic2000,diazTopologyOptimizationMethod2010}, the level set method \cite{luTopologyOptimizationAcoustic2013a}, the parametric level set method \cite{luoLevelSetbasedParameterization2008}, and BESO \cite{huangTopologicalDesignMicrostructures2011a}, among others. The level set approaches have the benefit of being flexible enough to consider various objective functions and mechanical models, including the design of auxetic metamaterials \cite{luoLevelSetbasedParameterization2008}. In contrast, the SIMP technique is the most popular approach because of its straightforward conceptualization and application.

Topology optimization for metamaterials relies on iterative finite element analysis and sensitivity analysis. Due to the numerous iterations and large numbers of elements, this process leads to substantial computational costs. Several strategies have been developed to alleviate this computational expense, which includes the development of efficient and fast solvers and the reduction of design variables \cite{amir2010efficient, wang2007large, borrvall2001large}.

Adopting a novel perspective, it can be hypothesized that obtaining a high-resolution (HR) optimized topology from an initial low-resolution (LR) optimized topology can significantly decrease computational cost. However, the relationship may remain complex and hidden, creating challenges for traditional techniques in achieving accurate results \cite{wang2021deep}. The advancement of data processing and deep learning algorithms has given rise to potent tools for extracting complex relationships through a learning process \cite{lecun2015deep}. Certainly, deep learning models have demonstrated exceptional performance across a broad spectrum of fields, spanning fracture evolution prediction \cite{jiang2022auto}, the discovery of governing physics \cite{manikkan2023transfer}, natural language processing (NLP), and the layout optimization of vehicles \cite{gantovnik2006multi}. DL has found applications in numerous engineering challenges as well \cite{yan2023real,meng2017ultrasonic,wang2018model,linden2023neural}.

The super-resolution (SR) technique generates an HR image corresponding to its LR counterpart observed in the data \cite{park2003super}. SR techniques are commonly employed in image processing to address the constraints posed by LR images. They achieve this by creating HR images that closely resemble the actual ground truth images in terms of visual similarity. The result produced by topology optimization is depicted as a monochromatic pixel image. Elevating the resolution of topology from lower to higher levels is identified as a Super-Resolution (SR) problem. SR technique has practical applications in various fields \cite{greenspan2009super,li2015super}.

In contrast, efficient integration of deep convolutional neural networks (CNNs) into SR, instead of conventional SR methods, can substantially enhance reconstruction performance. Dong et al. \cite{dong2014learning} were the first to incorporate DL into the domain of SR by designing SR-CNN, which aimed to ensure the highest achievable restoration quality. Inspired by the concept of residual networks \cite{he2016deep}, Kim et al. \cite{kim2016accurate} introduced the concept of very-deep SR, which proved to be a highly precise approach by extending the network to 20 layers. Ledig et al. \cite{ledig2017photo} introduced a super-resolution generative adversarial network (SRGAN), a framework rooted in the principle of generative adversarial networks (GANs). The original residual block within SRGAN showcased its effectiveness in handling image classification and detection assignments. However, the direct application of the original residual blocks for the SR problem yielded suboptimal performance. Consequently, Lim et al. \cite{lim2017enhanced} fine-tuned the residual block, increased the scale of the model, and introduced the SR technique.

In the context of metamaterial design, the LR input data is in the form of an LR-optimized topology mapped to an HR-optimized topology. Utilizing SR techniques for metamaterial design offers several advantages. It minimizes the computational expenses needed to create HR topologies since the SR network can directly generate them from LR inputs without requiring computationally expensive simulations or optimization algorithms. Also, this is useful for metamaterials with complex geometries or requiring precise microstructure control. Finally, it improves the accuracy and reliability of the design process by generating high-quality designs optimized for specific performance metrics or other design requirements.

The algorithm employs the SIMP and energy-based homogenization approach to generate optimized design data. The effective constitutive parameters are calculated using the mutual energies of the elements. This framework minimizes or maximizes various objective functions with one constraint. The microstructures are assumed to comprise periodic base cells. The mechanical characteristics of structures at the microscale are related to the material's general macroscale characteristics using the homogenization theory. This theory is used to compute the sensitivity numbers for each base cell. 

The current study develops an EDSR-based neural network framework to predict optimized HR metamaterial topologies. The approach establishes the mapping between LR and HR topologies, enabling efficient topology optimization across different design resolutions. The trained EDSR model is applied to optimize mechanical metamaterials for the maximization of bulk, elastic, and shear moduli, and the minimization of Poisson’s ratio. The major scientific contributions and highlights of this study are summarized as follows:
\begin{itemize}
    \item Developed a \textbf{deep learning framework} that integrates deep super-resolution with SIMP-based topology optimization to efficiently design high-resolution mechanical metamaterials.
    \item Introduced an \textbf{enhanced low-to-high-resolution mapping strategy} that enables the prediction of optimized high-resolution and \textbf{manufacturable topologies} from low-resolution datasets, significantly reducing computational requirements.
    \item Demonstrated that the proposed approach achieves \textbf{high predictive accuracy} across various design objectives, including maximization of bulk, shear, and equivalent elastic moduli, and minimization of Poisson’s ratio (auxetic metamaterials).
    \item Achieved high predictive accuracy and up to \textbf{93–95$\%$ reduction in computational cost} compared to conventional SIMP-based TO.
    \item Validated the proposed framework using comprehensive \textbf{quantitative evaluation metrics}, confirming its accuracy, efficiency, and scalability for high-resolution metamaterial design.
\end{itemize}

Following is the organization of this paper. The SIMP-based topology optimization (TO) approach used to generate metamaterial microstructures is briefly described in Section \ref{sec:meta_formulation}. The details of the proposed framework, including the convolutional neural network (CNN) architecture and the training process, are presented in Section \ref{sec:approach}. Extensive numerical investigations are reported in Section \ref{sec:NE}, where the variation of optimized geometries and corresponding mechanical properties with different parameters is analyzed, along with validation against benchmark problems. Finally, Section \ref{sec:conclusion} summarizes the key findings and conclusions of the study.

\section{SIMP-based metamaterial topology optimization}
\label{sec:meta_formulation}
The energy-based homogenization approach is used in this study to estimate the macroscopic equivalent elastic characteristics of metamaterials. We utilize SIMP-based metamaterial topology optimization to generate data for EDSR. 

\subsection{Homogenization} 
\begin{figure}
    \centering
    \includegraphics{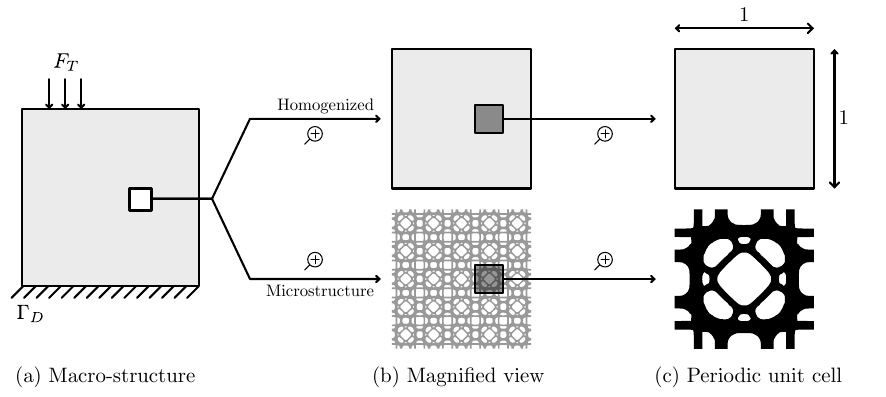}
    \caption{Illustration of a material constituted by periodically arranged microstructures.}
\label{fig:metamaterials}
\end{figure}

In the 1970s, the homogenization theory emerged as an alternate way to laboratory investigation for determining effective material characteristics \cite{hassaniReviewHomogenizationTopology1998}. The energy-based homogenization approach is used in this study to estimate the macroscopic equivalent elastic characteristics of metamaterials \cite{xiaDesignMaterialsUsing2015}. Considering a unit cell arranged periodically characterizes the behavior of the metamaterial, as shown in Fig.~\ref{fig:metamaterials}. Let $\boldsymbol{\sigma}_{ij}$ and $\boldsymbol{\epsilon}_{ij}$ denote the stress and strain, respectively, within the unit cell, and let $\overline{\boldsymbol{\sigma}}_{ij}$ and $\overline{\boldsymbol{\epsilon}}_{ij}$ denote the stress and strain, respectively, within the homogenized metamaterial. The effective elasticity matrix represents the behavior of the metamaterial, directly dependent on the unit cell's architecture and the base material's elastic properties. The elasticity matrix of the base material is denoted by $\boldsymbol{C}_{ijkl}$. $\overline{\boldsymbol{C}}_{ijkl}$ represents the elasticity matrix of the metamaterial. The constitutive relation of base material and the homogenized metamaterial is:

\begin{equation} \label{eq:constitutive relation base}
    \boldsymbol{\sigma}_{i j}=\boldsymbol{C}_{ijkl} \boldsymbol{\varepsilon}_{kl}
\end{equation}
\begin{equation} \label{eq:constitutive relation}
    \overline{\boldsymbol{\sigma}}_{i j}=\overline{\boldsymbol{C}}_{ijkl}\overline{\boldsymbol{\epsilon}}_{k l}
\end{equation}

For the 2D orthotropic material, we can write $\overline{\boldsymbol{C}}_{ijkl}$ as:

\begin{equation} {\label{eq:constitutive_2D}}
    \overline{\boldsymbol{C}}_{ijkl}=\left[\begin{array}{ccc}
    \overline{C}_{1111} & \overline{C}_{1122} & 0 \\
    \overline{C}_{1122} & \overline{C}_{2222} & 0 \\
    0 & 0 & \overline{C}_{1212}
    \end{array}\right]=\frac{E}{1-\bar{\nu}^{2}}\left[\begin{array}{ccc}
    1 & \bar{\nu} & 0 \\
    \bar{\nu} & 1 & 0 \\
    0 & 0 & \frac{(1-\bar{\nu})}{2}
    \end{array}\right]
\end{equation}

where 
$\bar{\nu}$ is the Poisson's ratio of unit cells. Let $V$ denote the volume of the unit cell. The overall strain energy of the unit cell, taking the parameters of the base material into account, is given by:
\begin{equation}
\begin{split}
\label{eq:energy_unit}
        U =\frac{1}{2} \int_{V} \boldsymbol{\sigma}_{i j} {\boldsymbol{\epsilon}}_{k l} dV
         = \frac{1}{2} \int_{V} \boldsymbol{C}_{ijkl} {\boldsymbol{\epsilon}}_{k l} {\boldsymbol{\epsilon}}_{i j} dV
\end{split}
\end{equation}

The overall strain energy, when metamaterial properties are taken into account, is given as:

\begin{equation} \label{eq:strain}
\begin{split}
        \overline{U} =\frac{1}{2} {V} \overline{\boldsymbol{\sigma}}_{i j} \overline{{\boldsymbol{\epsilon}}}_{k l}
         = \frac{1}{2} {V} \overline{{\boldsymbol{\epsilon}}}_{i j} \overline{\boldsymbol{C}}_{i j k l} \overline{{\boldsymbol{\epsilon}}}_{k l}
\end{split}
\end{equation}

According to the strain-energy based method \cite{zhangUsingStrainEnergybased2007}, for elastic problems, the total strain energy of the unit cell ($U$) is equal to the total strain energy of the unit cell considering it as a homogeneous medium ($\overline{U}$), i.e.

\begin{equation} \label{eq:strain_energy}
    U = \overline{U}
\end{equation}

Substituting Eq.~\ref{eq:constitutive relation}  into Eq.~\ref{eq:strain} and applying the condition of Eq.~\ref{eq:strain_energy} , we get:

\begin{equation} \label{eq:homo_elastic_tensor}
U=\frac{1}{2}V\left[\begin{array}{c}
\bar{\varepsilon}_{11} \\
\bar{\varepsilon}_{22} \\
\bar{\varepsilon}_{12}
\end{array}\right]^T\left[\begin{array}{ccc}
\overline{C}_{1111} & \overline{C}_{1122} & 0 \\
\overline{C}_{1122} & \overline{C}_{2222} & 0 \\
0 & 0 & \overline{C}_{1212}
\end{array}\right]\left[\begin{array}{c}
\bar{\varepsilon}_{11} \\
\bar{\varepsilon}_{22} \\
\bar{\varepsilon}_{12}
\end{array}\right]
\end{equation}

Using Eq.~\ref{eq:homo_elastic_tensor}, the coefficients of the homogenized elastic tensors ($\overline{C}_{1111}$, $\overline{C}_{1122}$, and $ \overline{C}_{2222}$) can be calculated. 

\begin{figure}
    \centering
    \includegraphics{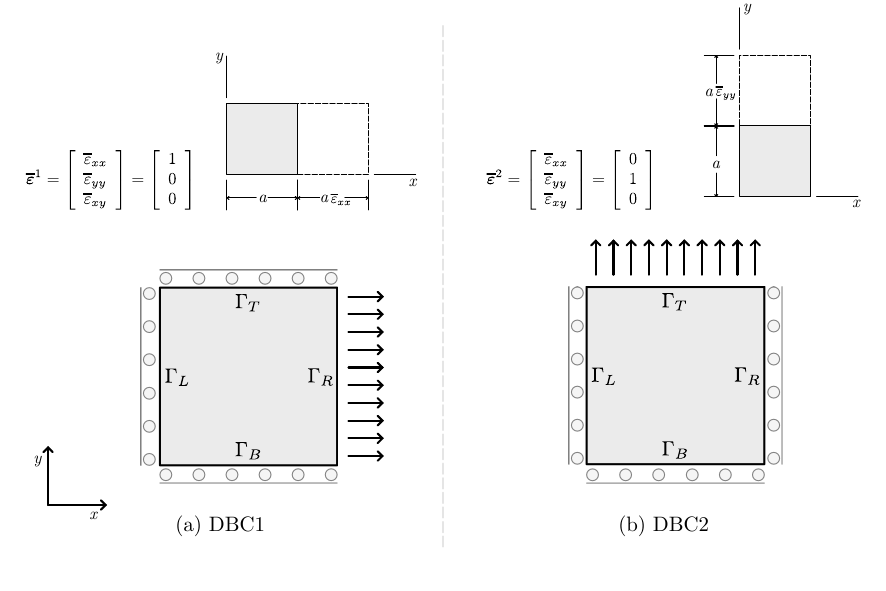}
    \caption{Dirichlet boundary conditions (DBC's): (a)DBC1$:[u_{x}=0 \text{ on }  \Gamma_L, u_{y}=0 \text{ on }  \Gamma_T, u_{y}=0 \text{ on }  \Gamma_B, u_{x}=1 \text{ on }  \Gamma_R]$.  (b) DBC2$:[u_{x}=0 \text{ on }  \Gamma_L, u_{y}=1 \text{ on }  \Gamma_T, u_{y}=0 \text{ on }  \Gamma_B, u_{x}=0 \text{ on }  \Gamma_R]$.}
    \label{fig:boundary}
\end{figure}

Considering a unit cell $\Omega \subset \mathbb{R}^2$ as shown in Fig.~\ref{fig:boundary}, having boundary $\Gamma$, with $\Gamma_L,\ \Gamma_R,\ \Gamma_T,\ \text{and }\Gamma_B$ representing the left, right, top and bottom boundaries, respectively. The uniform strain fields  ($\overline{{\boldsymbol{\varepsilon}}}^1$, $\overline{{\boldsymbol{\varepsilon}}}^2$) can be imposed by applying the Dirichlet boundary conditions (DBCs) as represented in Fig.~\ref{fig:boundary}(a) and Fig.~\ref{fig:boundary}(b) for case-1 and case-2 respectively, with the application of the following unit strain fields:

\begin{equation} \label{eq:unit_strain}
\overline{{\boldsymbol{\varepsilon}}}^1=\left\{\begin{array}{l}
1 \\
0 \\
0
\end{array}\right\}, \quad \overline{{\boldsymbol{\varepsilon}}}^2=\left\{\begin{array}{l}
0 \\
1 \\
0
\end{array}\right\}
\end{equation}

To evaluate $\overline{C}_{1111}$, we consider $\overline{{\boldsymbol{\varepsilon}}}^1$ strain field from Eq.~\ref{eq:unit_strain}, substitute it in Eq.~\ref{eq:homo_elastic_tensor}
\begin{equation}
    \begin{aligned}
    & \therefore U_{1111}=\frac{1}{2} V\left[\begin{array}{l}1 \\0 \\0\end{array}\right]^{\top}\left[\begin{array}{ccc}\overline{C}_{1111} & \overline{C}_{1122} & 0 \\\overline{C}_{1122} & \overline{C}_{2222} & 0 \\0 & 0 & \overline{C}_{1212}\end{array}\right]\left[\begin{array}{l}1 \\0 \\0\end{array}\right]
    \end{aligned}
\end{equation}

\begin{equation}
\begin{aligned}
& \Rightarrow \quad \overline{C}_{1111}=\frac{2 U_{1111}}{V}
\end{aligned}
\end{equation}

To evaluate $\overline{C}_{1122}$, we can consider the $\overline{{\boldsymbol{\varepsilon}}}^1$ and $\overline{{\boldsymbol{\varepsilon}}}^2$ strain fields from Eq.~\ref{eq:unit_strain} together and  substitute it in Eq.~\ref{eq:homo_elastic_tensor}:
\begin{equation}
\begin{aligned}
U_{1122} &=\frac{1}{2} V\left[\begin{array}{l}
1 \\
0 \\
0
\end{array}\right]^{T}\left[\begin{array}{ccc}
\overline{C}_{1111} & \overline{C}_{1122} & 0 \\
\overline{C}_{1122} & \overline{C}_{2222} & 0 \\
0 & 0 & \overline{C}_{1212}
\end{array}\right]\left[\begin{array}{l}
0 \\
1 \\
0
\end{array}\right] 
\end{aligned}
\end{equation}

\begin{equation}
\begin{aligned}
& \Rightarrow \quad \overline{C}_{1122}=\frac{2 U_{1122}}{V}
\end{aligned}
\end{equation}

\subsection{Optimization approach}
\label{sec:TO_formulation}

The unit cell's architecture influences the properties of metamaterials. A suitable unit cell architecture can be designed to enhance the material properties using topology optimization (TO) in conjunction with homogenization, assuming constant base material properties. TO seeks the optimal material distribution within a design domain, typically expressed as a density field ($\phi \in \mathbb{R}^2$), while adhering to specified design goals and constraints. In the density-based method (DBM), a discontinuous density field is assigned to finite elements, with values ranging from 0 to 1 to indicate the material proportion in each region. The optimization problem for microstructure TO can be described as follows:
\begin{equation}  \label{eq:comp}
    \begin{aligned}
     \underset{\phi}{\operatorname{minimize}}&\quad f(\phi)  \\
     \text {subjected to}&\quad g(\phi) : \frac{\int \phi \mathrm{d} \Omega}{\mathcal{V}_f \int 1\mathrm{d} \Omega} -1\leq 0 \\
    &\quad  0 \leq \phi_e \leq 1, e=1, \ldots, N.
    \end{aligned}
\end{equation}
In the equation above, $\mathcal{V}_f$ represents the desired material volume fraction. The objective function $f(\phi)$ can be defined as follows: 
\begin{equation}  \label{eq:objective_function}
    f(\phi)=-\eta(\phi_e)c
\end{equation}
where $\eta(\phi_e)$ is the SIMP penalization function. This function employs a power law to link the density variable with material properties, discouraging material presence in an element either to a full (1) or none (0). The $c$ can be further defined according to desired properties as \cite{xiaDesignMaterialsUsing2015}: 

\begin{itemize}
  \item Design of maximum bulk modulus:
\end{itemize}
\begin{equation}\label{eq:BM}
    c=-\left(\overline{C}_{1111}+\overline{C}_{1122}+\overline{C}_{2211}+\overline{C}_{2222}\right)
\end{equation}

\begin{itemize}
  \item Design of maximum shear modulus:
\end{itemize}
\begin{equation} \label{eq:SM}
c=-\overline{C}_{1212}
\end{equation}

\begin{itemize}
  \item Design of maximum equivalent elastic modulus:
\end{itemize}
\begin{equation} \label{eq:EM}
c=-\frac{1}{2}\left(\overline{C}_{1111}+\overline{C}_{2222}\right)
\end{equation}

\begin{itemize}
  \item Design of negative Poisson's ratio:
\end{itemize}
\begin{equation} \label{eq:NPR}
c=\overline{C}_{1122}-\beta^l(\overline{C}_{1111}+\overline{C}_{2222})
\end{equation}

where $\overline{C}_{ijkl}$ is the homogenized stiffness tensor, $\beta$ is the weighting factor and $l$ is the iteration number. Following the modified SIMP approach, we can define the SIMP penalization function $\eta(\phi_e)$ as described in \cite{sigmundMorphologybasedBlackWhite2007}:
\begin{equation} \label{eq:simp}
    \eta(\phi_e)=E_{\min }+\left(\phi_e\right)^p\left(E_0-E_{\min }\right), \quad \phi_e \in[0,1]
\end{equation}
In Eq.~\ref{eq:simp}, $p$ serves as the penalization parameter, and we denote Young's modulus of the solid material as $E_0$. Meanwhile, Young's modulus of the Ersatz material is denoted as $E_{\min}$, and it's set to be $1e-9$ times $E_{0}$. The function $\phi$ is a piecewise discontinuous function with a constant value within a single element and exhibits discontinuities at element boundaries. Consequently, we classify the function $\phi$ as belonging to the space comprised of discontinuous piecewise Lagrange polynomials, which can be summarized as follows:
 \begin{equation}
 \label{eq:dg}
     D=\left\{\phi \in L^2(\Omega):\left.\phi\right|_T \in L_\mathcal{P}(T) \forall T \in \mathcal{T} \text { and } \mathcal{P}=0\right\}
 \end{equation}
In Eq.~\ref{eq:dg}, $\mathcal{T}$ represents all cells $T$ within the finite element mesh, and $L_\mathcal{P}(T)$ denotes a function space comprising discontinuous Lagrange elements of degree $\mathcal{P}$. We opt for $\mathcal{P}=0$ to achieve a piecewise constant function.

\subsection{Sensitivity analysis and mesh independency filter} 
To conduct the sensitivity analysis, we compute the derivative of the objective function Eq.~\ref{eq:objective_function} with respect to the design variables, as follows:

\begin{equation}  \label{eq:sensitivity}
    \frac{\partial f}{\partial \phi_e}= -p \phi_e^{p-1}\left(E_0-E_{\min }\right) c
\end{equation}

\begin{figure}
    \centering
    \includegraphics{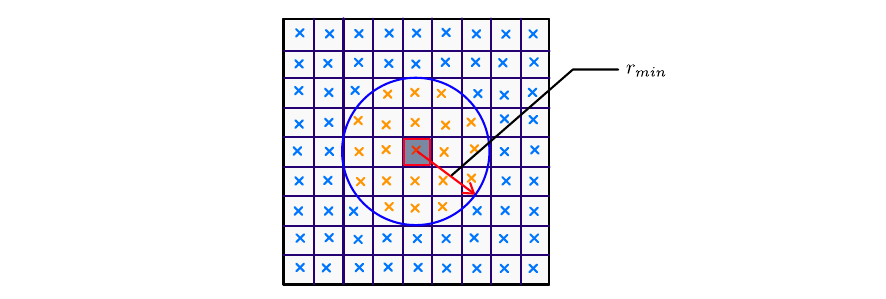}
    \caption{Cone filter. The cone filter functions by performing a weighted averaging of parameter values from neighboring elements or cells located within a defined radius. (Top) Elements denoted as $ele_b$ within the vicinity of $ele_a$ are taken into account for the density filtering process. (Bottom) The weight factor ($W_b$) varies as the distance from $ele_a$ increases. It is important to observe that the weight factor becomes zero beyond the filter zone indicated by $r_{min}$.}
    \label{fig:filter}
\end{figure}

Common problems encountered in topology optimization (TO), such as mesh sensitivity and checker-boarding, have led to the development of various methods to mitigate these issues. These methods include mesh independence filters, which encompass sensitivity filters and density filters \cite{sigmundNumericalInstabilitiesTopology1998, andreassenEfficientTopologyOptimization2011}. In our research, we applied the sensitivity filter proposed by \cite{sigmundMorphologybasedBlackWhite2007}, which involves averaging sensitivities over a nearby region of the current element (refer Fig.~\ref{fig:filter}). This process modifies the sensitivities $\frac{\partial f}{\partial \phi_{e}}$ as follows:
\begin{equation}  \label{eq:sensitivity_filter}
\frac{\widehat{\partial f}}{\partial \phi_{a}}=\frac{1}{\phi_a \sum_{b=1}^N W_b} \sum_{b=1}^N W_b \phi_{i} \frac{\partial f}{\partial \phi_{i}}
\end{equation}
The convolution operator, denoted as $W_b$, is defined as follows:
\begin{equation}\label{eq:weight}
    \left.W_b=r_{\min }-\operatorname{dist}(b, a),\left\{b \in \mathbb{N} \mid \operatorname{dist}(b, a) \leq r_{\min }\right)\right\}, b=1, \ldots, N
\end{equation}
In this context, `$a$' and `$b$' represent distinct elements, with `dist $(b, a)$' signifying the distance between their midpoints. The convolution operator, denoted as $W_b$, linearly decreases as the distance from element `$a$' increases. Any region outside the circle with a filter radius of $(r_{min})$ has a weight factor of zero.

\subsection{Optimality criteria method}
To address the optimization problem described in Eq.~\ref{eq:comp}, we utilize the standard optimality criteria (OC) method. We update the design variables by following the gradient-based updating scheme introduced by \cite{bendsoeOptimizationStructuralTopology1995}, as shown below:

\begin{equation} \label{eq:oc}
\phi_{e}^{\text {new }} = \begin{cases}
\max \left(0, \phi_{e}-m\right), & \text { if } \phi_{e} B_{e}^{\eta} \leq \max \left(0, \phi_{e}-m\right), \\
\min \left(0, \phi_{e}+m\right), & \text { if } \phi_{e} B_{e}^{\eta} \geq \min \left(0, \phi_{e}+m\right), \\
\phi_{e} B_{e}^{\eta}, & \text {otherwise} \\
\end{cases}
\end{equation}
where $B_{e}$ can be calculated as:

\begin{equation}  \label{eq:be}
    B_{e}=\frac{-\frac{\widehat{\partial f}}{\partial \phi_{e}}}{\lambda \frac{\partial V}{\partial \phi_{e}}}
\end{equation}
where, the Lagrange multiplier $\lambda$ is determined using the bisection algorithm.

\subsection{Algorithm implementation}
The steps in data generation for metamaterials TO used for the study are shown in Algorithm~\ref{alg:cap}. We began by defining the design domain and material properties of the metamaterial. The density variable was initialized, and various control parameters, such as volume fraction, penalization power, filter radius, mesh size, initial design, and the objective function, were specified based on the data generation problem.

Next, we vary one parameter at a time while keeping others constant. We solve the equilibrium equation within each iteration and compute homogenized elasticity tensors. These homogenized elasticity tensors are used to frame the objective function for specific mechanical properties. The sensitivity is defined as the differentiation of objective function with respect to the density variable. After applying the sensitivity filter, we solve the optimization problem using OC. The algorithm checks for stopping criteria and gets terminated when the criteria are met; otherwise, it moves to the next iteration. This way, data generation has been done. The EDSR neural network receives LR topologies derived from generated data as input, and HR topologies are employed to compute data loss during the training process of EDSR.

\begin{algorithm}
\caption{Methodology for generating data from metamaterials topology optimization}\label{alg:cap}
\begin{algorithmic}[1]
\State Define the design space and material properties of the metamaterial.
\State Initialize the density variable ($\phi_{e}$) and mesh size ($nel_x \times nel_y$).
\State Define topological control parameters: (i) $\mathcal{V}_f$, (ii) $p$, (iii) $r_{min}$ and (iv) objective function.
\State Vary one parameter at a time and set all other parameters constant.
\State Define maxIter, set $i$ = 0  		
\For {$i = 0$ to maxIter}  
\State Compute the objective function $f(\phi)$ from Eq.~\ref{eq:comp} and Sensitivity $\frac{\partial f}{\partial \phi_e}$ from Eq.~\ref{eq:sensitivity}.
\State Apply sensitivity filter using Eq.~\ref{eq:sensitivity_filter}.
\State Solve the optimization problem using OC with Eq.~\ref{eq:oc}. 
\If{stopping criteria is achieved}
\State stop
\EndIf
\EndFor
\end{algorithmic}
\end{algorithm}

\section{Deep super-resolution framework for metamaterial design}
\label{sec:approach}
\begin{figure}
    \centering
    \includegraphics{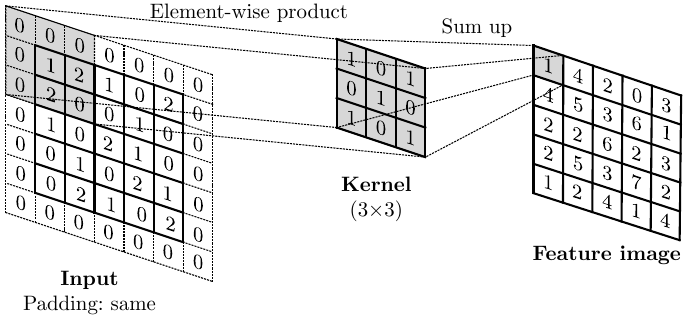}
    \caption{Illustrating a convolution operation instance featuring a $3 \times 3$ kernel size, same padding, and a stride of one. In this procedure, the kernel moves across the input tensor, and at each location, it performs an element-wise multiplication with the input tensor. The results are added together to obtain the output value at the respective position in the output tensor, commonly known as a feature map.}
    \label{fig:CNN_layer}
\end{figure}

Convolutional Neural Network (CNN) is a special category of artificial neural networks primarily designed for image processing, renowned for its ability to systematically generate abstract representations through convolution operations \cite{gu2018recent}. CNN exploits the inherent two-dimensional (2D) image structure and harnesses shared correlations among adjacent pixels. The core strength of CNN lies in convolutional filters (kernel), which can be expressed as weight matrices. The illustration of a convolution operation instance featuring a $3 \times 3$ kernel size, same padding, and a stride of one is shown in Fig.~\ref{fig:CNN_layer}. These kernels traverse the input image, producing a feature image as the resulting output.

\begin{figure}
    \centering
    \includegraphics{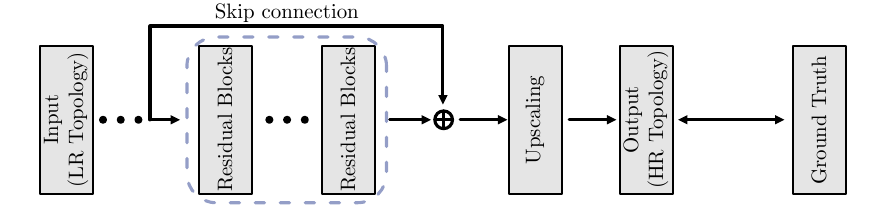}
    \caption{Illustrating the detailed flow of an enhanced deep super-resolution (EDSR) and its key component, it shows the step-by-step process of enhancing image resolution.}
    \label{fig:conceptual}
\end{figure}

EDSR utilizes CNN as a mapping function to transform LR images into HR images. Fig.~\ref{fig:conceptual} illustrates the fundamental architecture of the EDSR technique. The EDSR architecture has three main components that are responsible for better model performance. These components are deep neural networks (residual blocks), upsampling operation (convolutional and sigmoidal operation), and skip connections. Deep neural networks exhibit greater accuracy compared to shallow neural networks. Recognizing the significance of low-level features in the overall performance, an extra skip connection is added to combine feature images from different levels. At last, the process of upsampling increases the image resolution. 

\begin{figure}
    \centering
    \includegraphics{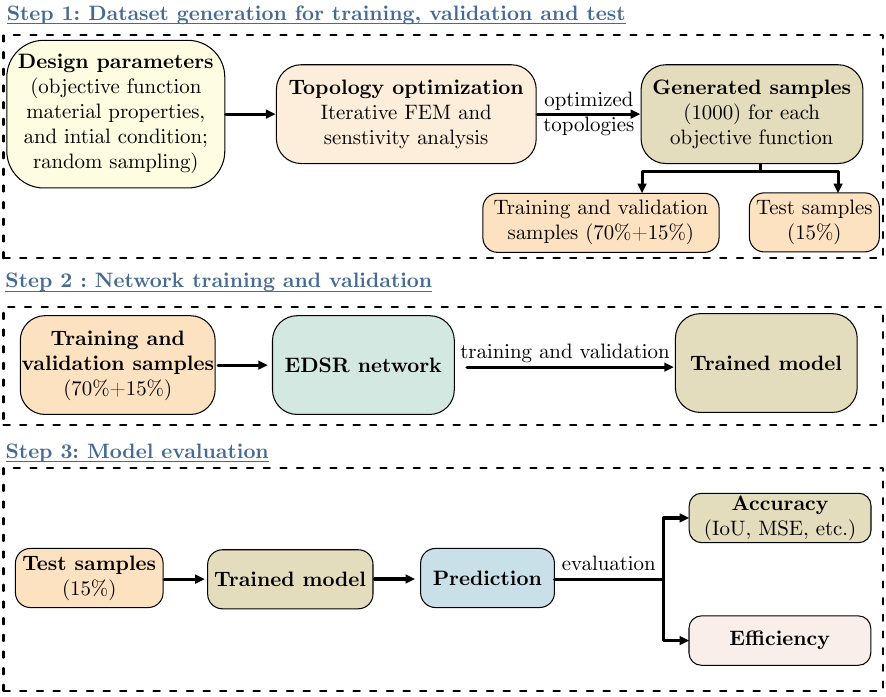}
    \caption{Flowchart of an EDSR-based approach for metamaterial TO. Illustrating the iterative workflow of an Enhanced Deep Super-Resolution (EDSR) based approach applied to metamaterial topology optimization, encompassing parameter initialization, EDSR network training, and model evaluation.}
    \label{fig:approach}
\end{figure}

A graphical representation depicting the proposed approach is shown in Fig.~\ref{fig:approach}. In Step 1, we generate the dataset using SIMP-based TO to validate, train, and test the model. We initialize with design parameters: objective function, material properties, initial design, and topological control parameters. Random sampling is carried out to choose the design parameters. Next, we perform SIMP-based TO of metamaterial for chosen parameters by following the steps outlined in Algorithm~\ref{alg:cap}. The generated dataset includes 1000 samples for each design objective. Every sample comprises both low- and high-resolution topologies. We split the dataset into three parts, 70\%, 15\%, and 15\%, for training, validation, and testing, respectively. Step 2 involves training and designing an EDSR network utilizing the TensorFlow library of Python \cite{developers2022tensorflow}. Section~\ref{sec:network} describes the architectural details of the EDSR network and detailing of its building blocks. We train and validate our model using training and validation data described in Section~\ref{sec:training}. In the subsequent step (Step 3), we import the trained model's parameters to enable the prediction of our test samples. These predictions involve evaluating our proposed method's computational efficiency and accuracy by comparing them with the corresponding ground truths. Section~\ref{sec:evaluation} details the comparison metrics employed in our study. The initial investment of time required for sample preparation and network training constitutes a one-time, initial offline expenditure. After training, the model parameters are saved for future access, enabling the prediction of HR topologies under various design objectives without additional time-consuming steps.

\subsection{Dataset generation and preprocessing} 
\label{sec:data_gen}
\begin{figure}
    \centering
    \includegraphics{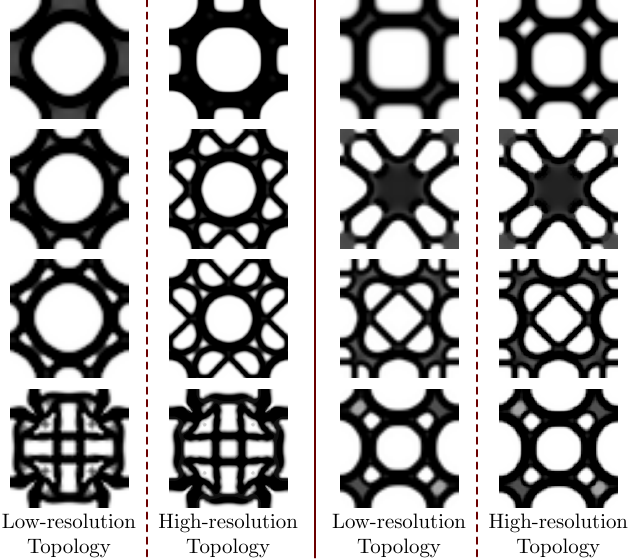}
    \caption{Generated dataset using SIMP-based TO for randomized parameters (each LR optimized topology corresponds to their HR optimized topology).}
    \label{fig:gen_data}
\end{figure}

To construct the dataset for the neural network, we generate a total of 4000 optimized topologies (1000 for each objective function). SIMP-based TO of metamaterial is used to generate the datasets. The samples are split into training, testing, and validation sets, with each set being independent of the others, following a ratio of 70\% for training, 15\% for testing, and 15\% for validation. The design domain is divided into LR $(48\times48)$ and HR $(192\times192)$. Fig.~\ref{fig:gen_data} depicts both low and high-resolution generated data samples (topologies) using SIMP-based TO for the various design objectives and randomized parameters. A random parameter sampling technique is employed to create the training, testing, and validation datasets. Following are the details of the randomized parameters:

\begin{itemize}
    \item \textbf{Design objective:} maximization of bulk modulus, maximization of elastic modulus, maximization of shear modulus, and minimization of Poisson's ratio.
    \item \textbf{Mesh size:} $48\times48$ and $192\times192$.
    \item \textbf{Volume fraction:} $0.2 - 0.7$ (uniformly distributed).
    \item \textbf{Penalization parameter:} $3 - 7.5$ (uniformly distributed).
    \item \textbf{Filter radius:} $0.02 - 0.07$ units (uniformly distributed).
\end{itemize} 

\subsection{EDSR network architecture}
\label{sec:network}
\begin{figure}
\centering
\includegraphics{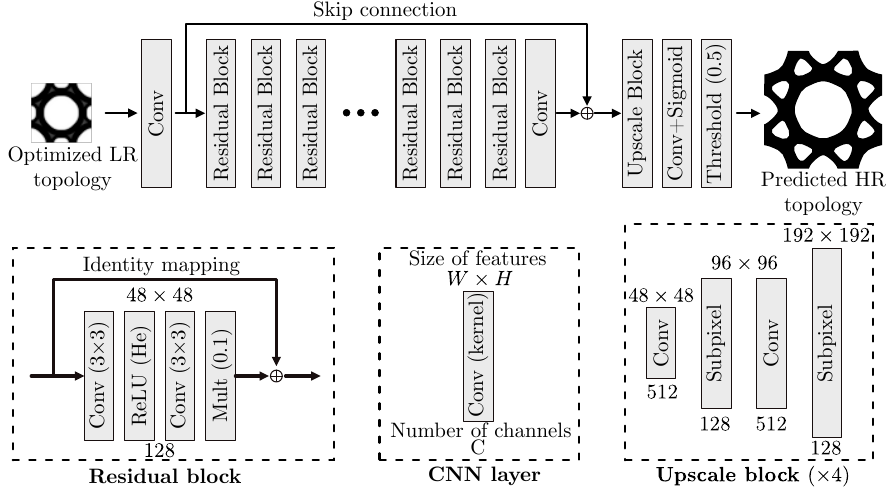}
\caption{Architectural design of a neural network intended for high-resolution topology generation from low-resolution input data. {\normalfont ``Conv''} represents the convolution layer, {\normalfont ``Multi''} denotes the convolution operation with a residual factor of 0.1, {\normalfont ``ReLU''} signifies the rectified linear unit, and {\normalfont ``CNN''} stands for the convolutional neural network.}
\label{fig:architecture}
\end{figure}

The complex EDSR network architecture employed for our study is illustrated in Fig.~\ref{fig:architecture}. This network is trained on pre-processed LR and HR topologies. EDSR network consists of the residual blocks that help to address the challenge of vanishing gradients and achieve deeper network architectures \cite{wang2021deep}. We have taken 12 residual blocks, with each individual block comprising two convolution layers utilizing convolution kernels of size $ 3 \times 3 $, along with rectified linear unit (\texttt{ReLU}) activation function \cite{nair2010rectified}. We have increased the number of feature channels to 128 of the convolution layer. Incorporating 128 feature channels and 12 residual blocks results in a substantial increase in model size, thereby enabling the extraction of more intricate features. Nonetheless, this enhancement introduced numerical instability during the training process. A residual scaling factor of 0.1 was employed to address this stability issue.

To attain enhanced performance, it is recommended to initialize parameters for the \texttt{ReLU} using the \texttt{He} initialization method \cite{he2015delving}. It mitigates the vanishing and exploding gradient problem during training by initializing weights in a specific way. When using \texttt{ReLU} as activation functions, \texttt{He} initialization sets the initial weights based on a Gaussian distribution. This technique promotes stable and faster training, making it a popular choice for deep networks with \texttt{ReLU}.

In the upscaling block, two consecutive $2\times$ upscaling layers were used to augment the feature size and produce an HR topology. In the final step, we utilized a convolutional layer, followed by a sigmoid activation function \cite{mitchell1997machine}. This configuration confines the output within the 0 to 1 range, yielding the black-and-white topology. Finally, a threshold projection has been applied to deblur the region in the topologies, with a threshold value set at 0.5. We ultimately determine the network architecture and fine-tuned hyper-parameters through trial and error. These hyper-parameters include factors such as the filter size, the quantity of residual blocks and layers, the choice of the optimizer, and the learning rate.

\subsubsection{Residual block design}
In the EDSR model, residual blocks are critical in effectively recovering HR topologies from LR input topology. They are fundamental building blocks, particularly in residual neural networks \cite{he2016deep}. The key concept behind the residual block is residual learning, which aims to tackle the degradation problem encountered in very deep neural networks. In conventional deep networks, each layer seeks to learn the mapping from its input to its output directly. However, as the network depth increases, it becomes challenging for the network to learn the optimal mapping due to vanishing or exploding gradients \cite{he2016deep}. This problem inhibits the ability of deep networks to capture complex details in the input LR topologies.

Residual learning tackles this problem by incorporating skip connections that skip one or more layers. These skip connections empower the network to learn residual functions directly, capturing the disparities between the LR and HR topologies \cite{he2016deep}. By doing so, residual blocks facilitate the learning of fine-grained details crucial for increasing LR topologies' resolution. Fig.~\ref{fig:architecture} illustrates a standard residual block comprising two convolutional layers, followed by the activation function, \texttt{ReLU}. The skip connection provides a shortcut for gradient propagation, mitigating the issue of vanishing gradients and enabling the network to effectively learn the residuals.

\subsubsection{Upscaling module}
The primary purpose of upscale blocks is to upsample the LR topologies to match the desired HR topologies. Upscale blocks typically employ interpolation to upscale the feature maps spatially. The most commonly used methods for spatial upscaling include bi-linear interpolation, nearest neighbor interpolation, or more advanced techniques such as transposed convolution or sub-pixel convolution \cite{shi2016real}. Bi-linear interpolation is a simple yet effective method used in upscale blocks \cite{shi2016real}. It calculates new pixel values by taking weighted averages of neighboring pixels in the low-resolution feature map. Nearest neighbor interpolation, on the other hand, assigns the value of the nearest pixel to each new pixel in the upscaled feature map. While these methods are computationally efficient, they may not capture intricate details and smoothness in the upscaled image. 

In more advanced upscale blocks, convolutional operations are used to perform the spatial upscaling. Another technique is sub-pixel convolution, which rearranges the channels of the LR feature maps to form HR feature maps. This rearrangement effectively increases the spatial resolution of the feature maps. To serve our purpose, we utilize the sub-pixel convolution technique. We combine the upscale block with the convolutional operation to learn the upsampling process and extract higher-level features \cite{shi2016real}. The combined operation helps to preserve important details and enhance the visual quality of upscaled topologies.

\subsection{Training strategy and hyperparameter optimization}
\label{sec:training}
The training of a neural network is fundamentally formulated as an optimization problem, primarily focused on minimizing the disparity between the network's predicted topology and the corresponding target topology. In this paper, the EDSR neural network aims to align the predicted topology with a material distribution closely resembling the optimized high-resolution (HR) topology. The mathematical representation of the objective function is as follows \cite{wang2021deep}: 
\begin{equation}
    L(\textbf{y},h_\theta(\textbf{x}))=-E_{\textbf{x} \sim P_{data}}[\textbf{y}\log h_\theta(\textbf{x})+(1-\textbf{y})\log(1-h_\theta(\textbf{x})]
\end{equation}
where $L(\textbf{y},h_\theta(\textbf{x}))$ represents the network's loss function. In this context, $h_\theta(\textbf{x})$ corresponds to the probability density function of the samples, $\textbf{y}$ represents the label, and $\textbf{x}$ denotes the training sample. Within the scope of this article, $\textbf{x}$ pertains to the LR topology, $\textbf{y}$ is associated with the corresponding HR topology, and $h_\theta(\textbf{x})$ signifies the predicted topology generated by the neural network.
\begin{equation}
    \theta_t = \theta_{t-1}-\eta\cdot\nabla_\theta L(\textbf{y},h_\theta(\textbf{x}))
\end{equation}
The learning rate $\eta$ and gradient $\nabla_\theta L(\textbf{y},h_\theta(\textbf{x}))$ are the two fundamental components of the optimizer \cite{kingma2014adam}. The learning rate plays a pivotal role in shaping the neural network's parameter adjustments to align with the gradient loss, significantly influencing the entire learning process. Consequently, it emerges as a challenging hyper-parameter to select appropriately. Lowering the learning rate in gradient descent may slow the process, while using a higher learning rate can potentially result in overshooting the minimum \cite{kingma2014adam}. Therefore, the choice of optimizer and other hyper-parameters plays a role in avoiding local minima during the optimization process and ensuring its efficiency.

\subsubsection{Optimizer selection}
The optimizer's choice plays a significant role in hyper-parameter tuning for DL models \cite{sun2020optimization}. The optimizer directly affects the convergence speed, generalization performance, and overall effectiveness of the model, making it a crucial factor to consider. Optimizers employ different update rules and strategies for adjusting the model's parameters during training. Some optimizers may converge faster by taking larger steps in the parameter space, while others may use adaptive learning rates or momentum to navigate the optimization landscape more efficiently. The choice of optimizer can significantly impact the training time required to reach a satisfactory performance level.

Furthermore, different optimizers have strengths in handling specific problem characteristics. For instance, problems with sparse or unevenly scaled features may benefit from optimizers like AdaGrad or AdaDelta, which adapt the learning rate on a per-parameter basis \cite{sun2020optimization}. Conversely, problems associated with noisy or non-stationary data may find advantages in optimizers such as Adam, which dynamically adjust the learning rate depending on the first and second moments of the gradient \cite{kingma2014adam}. Understanding the problem's characteristics and choosing an optimizer that suits those characteristics improve the model's performance.

Another consideration is the robustness of optimizers to hyper-parameter choices. Some optimizers may be more sensitive to the initial learning rate or require careful tuning of additional hyper-parameters such as momentum or weight decay \cite{sun2020optimization}. Understanding the interplay between optimizers and other hyper-parameters is crucial for effective hyper-parameter tuning. To achieve the best performance, it is necessary to experiment and find the optimal combination of hyper-parameters, including the optimizer. Finally, the choice of optimizer can impact the model's generalization performance. Generalization refers to the ability of the model to perform well on unseen data. A well-selected optimizer can help the model converge to a global minimum or an optimal region that generalizes well to new examples. By selecting an optimizer that encourages better generalization, hyper-parameter tuning can lead to models performing well on the unseen and training data.

By considering all the aspects, we utilize the Adam optimizer to serve our purpose. The Adam optimizer is less prone to becoming trapped in local minima, and by fine-tuning hyper-parameters, it is possible to enhance model performance \cite{kingma2014adam}. The Adam optimizer, short for adaptive moment estimation, effectively addresses the challenge of determining an optimum learning rate. Adam, a gradient-based optimizer created for stochastic objective functions, dynamically refines the current learning rate by utilizing first and second-order gradient moment estimates. This adjustment is mathematically represented as \cite{kingma2014adam}
 \begin{equation}
     g_t = \nabla_\theta L(y,h_\theta(x))
 \end{equation}
 \begin{equation}
     b_t = \partial_1 \cdot b_{t-1}+(1-\partial_1) \cdot g_t
 \end{equation}
 \begin{equation}
     d_t = \partial_2 \cdot d_{t-1}+(1-\partial_2) \cdot g_t^2
 \end{equation}
 \begin{equation}
     \hat{b}_t=b_t/(1-\partial_1)
 \end{equation}
 \begin{equation}
     \hat{d}_t=d_t/(1-\partial_2)
 \end{equation}
 \begin{equation}
     \Delta_\theta = -\eta \frac{\hat{b}_t}{\sqrt{\hat{d}_t}+\epsilon}
 \end{equation}
 \begin{equation}
     \theta_t = \theta_{t-1}+\Delta_\theta
 \end{equation}
 where $b$ and $d$ represent the biased estimations of the first and second gradient moments. The variables $\partial_1$ and $\partial_2$ correspond to the exponential decay rates, and $\epsilon$ is a small increment. In this study, we have initialized the parameters as $\epsilon = 10^{-8}$, $\partial_1$ = $0.9$, $\partial_2$ = $0.999$ and $\eta$ = $10^{-4}$.

\begin{figure}
    \centering
    \includegraphics{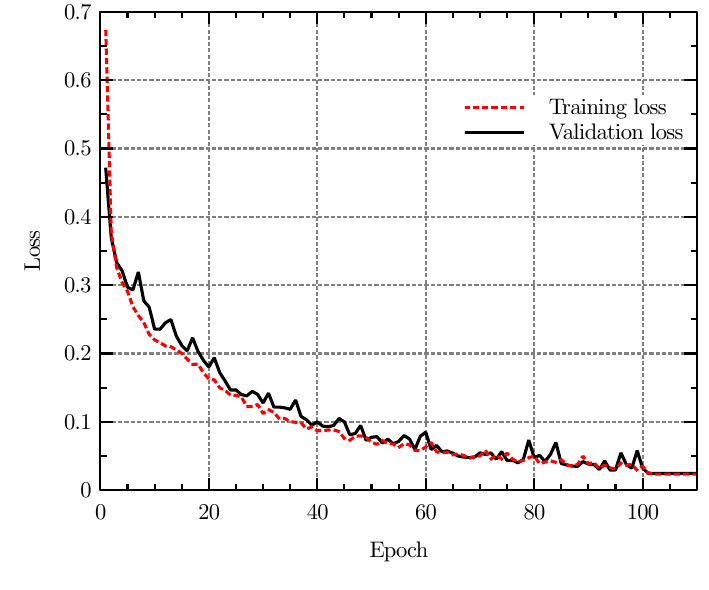}
    \caption{Loss evolution over time for training and validation sets. Training (red) and validation (black) loss over each epoch. It is inferred that training and validation errors exhibited concurrent decreases and converged to around 0.02 after 103 epochs.}
    \label{fig:loss_BM}
\end{figure}

Following the completion of one training epoch, validation of the neural network is performed. Fig.~\ref{fig:loss_BM} depicts the loss history observed during the training and validation for the case of maximization of bulk modulus. For this study, a batch size of 32 is utilized. Based on the loss curve, it is evident that the validation and training errors (0.0245 and 0.0233, respectively) exhibited concurrent decreases and converged at approximately 0.02 after 103 epochs.

\subsubsection{Other hyper-parameters settings}
\begin{itemize}
    \item \textbf{Learning rate:} It specifies the step size at which the model adjusts its parameters in optimization, relying on the loss function gradients. The learning rate substantially influences the convergence and accuracy of the model. Choosing an appropriate learning rate is crucial for successful model training. The training process may become unstable, and the loss function may fluctuate or diverge if the learning rate is too high. This can hinder convergence and prevent the model from reaching an optimal solution. Conversely, lowering the learning rate too much might lead to delayed convergence, requiring more training iterations to achieve the desired accuracy.

    \item \textbf{Batch size:} It specifies how many training samples are utilized in each forward and reverse pass during training. The choice of batch size can impact the speed at which the model converges to an optimal solution. Reduced batch sizes introduce additional noise into parameter updates, potentially accelerating convergence but causing more unpredictable progress. Increased batch sizes yield more consistent updates but might result in slower convergence.

    \item \textbf{Number of epoch:} Each epoch signifies a complete cycle through the training dataset, and the number of epochs determines the total iterations over the entire training dataset. Setting this hyperparameter correctly is crucial to avoid underfitting or overfitting. Underfitting occurs when the model hasn't seen enough data to learn the underlying patterns, resulting in poor performance. Conversely, overfitting occurs when the model learns the training data too well, including noise, and fails to generalize to unseen data. Properly setting the number of epochs can help balance these two extremes.

    \item \textbf{Depth and number of features:} In hyper-parameter tuning, finding the optimal model depth and feature count involves iterative testing and evaluation on training and validation datasets, considering metrics like accuracy or mean squared error. By varying these parameters, one seeks the balance that maximizes pattern capture without overfitting. The number of features determining the input space dimensionality is a critical hyperparameter. While more features can enhance learning, they may introduce noise and complexity, leading to overfitting. Consequently, it is critical to carefully pick the quantity of features to prevent these difficulties. Optimal depth and feature count are problem-specific; complex datasets may benefit from deeper architectures, while simpler ones may suffice with shallower models.
    
\end{itemize}

\subsection{Model evaluation metrics}
\label{sec:evaluation}
We employed a set of quantitative metrics to thoroughly evaluate the model's effectiveness. The volume fraction metric assesses the model's precision in predicting material distribution within the metamaterial structure. Simultaneously, the objective function error indicates how well the model optimizes metamaterial topologies. Pillow \cite{clark2015pillow} library of Python is utilized to convert the image into mesh to calculate the objective function of predicted topology. This mesh is further re-analyzed to compute each objective function using Eq.~\ref{eq:BM} for maximization of the bulk modulus, Eq.~\ref{eq:SM} for maximization of the shear modulus, Eq.~\ref{eq:NPR} for minimization of the Poisson's ratio, and Eq.~\ref{eq:EM} for maximization of the elastic modulus.

\begin{figure}
    \centering
    \includegraphics{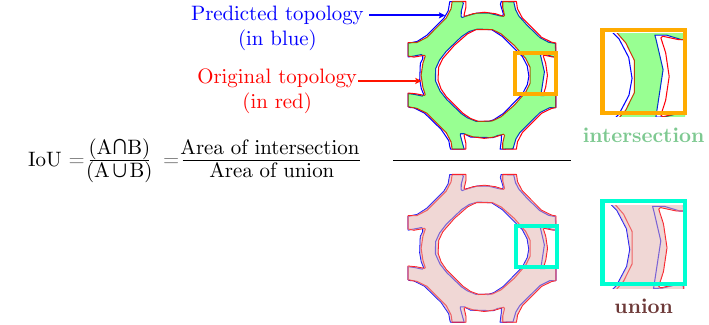}
    \caption{Intersection over union (IoU) representing spatial overlap where IoU = (area of intersected topologies) / (union of original and predicted topology). IoU measures the overlap between predicted bounding topologies and original topologies, with scores ranging from 0 to 1.}
    \label{fig:iou}
\end{figure}

Quantified by Mean Squared Error (MSE), Pixel-wise accuracy provided detailed insights into the alignment between the predicted topologies and the ground truth. Additionally, the intersection over union (IoU) metric facilitated the assessment of spatial agreement between predicted and ground truth topologies. IoU is calculated by dividing the intersection area between predicted and true values by the union of their respective areas, as illustrated in Fig.~\ref{fig:iou}. This metric offers a robust measure of accuracy and precision, providing valuable insights into the spatial correspondence and performance of the algorithms under scrutiny.

The error estimation process, integral to our methodology, is pivotal in comprehensively assessing the model's overall accuracy and efficiency. Following this evaluation, the results and discussion section unfolds, presenting a meticulous analysis that yields valuable insights into the robustness and potential limitations of the EDSR model within the nuanced domain of metamaterial topology optimization.

\section{Numerical experimentation}
\label{sec:NE}
In this section, we present a set of numerical experiments to evaluate the effectiveness and efficacy of our proposed method for designing mechanical metamaterials. The investigation is conducted using a single model trained on four distinct datasets, each corresponding to a specific metamaterial design objective: maximizing bulk modulus, elastic modulus, shear modulus, and minimizing Poisson's ratio. The 55-line Python code \cite{gupta55lineCodeLargescale2020}, developed using the open-source scientific computing platform FEniCS \cite{AutomatedSolutionDifferential, alnaesFEniCSProjectVersion2015}, has been used to generate data for this study. A comparative analysis is undertaken to assess the model's accuracy and efficiency, measured in terms of computational cost, against the conventional SIMP-based TO approach. Accuracy evaluation has been done in four aspects: pixel accuracy (quantified by mean squared error), volume fraction error, structural objective function error, and Intersection over Union. Furthermore, we systematically examine the adaptability and robustness of our approach by testing its efficacy on new, previously unseen data. All computations for data generation, training, validation, and testing of the model are conducted on a computational system equipped with an Intel Xeon 6230 processor and NVIDIA Quadro P2200 GPU.

\subsection{Maximizing the bulk modulus}
\label{sec:BM}
In the first study, we assess the performance of our approach in designing metamaterials to maximize bulk modulus. The LR and HR topologies, for mesh sizes of $48 \times 48$ and $192 \times 192$, are generated using SIMP-based topology optimization (as described in Section~\ref{sec:data_gen}). These shapes are generated using Eq.~\ref{eq:BM} for maximizing the bulk modulus as an objective function and are subsequently employed for training and testing the model. A dataset of 1,000 samples is generated by varying the volume fraction, filter radius, and penalization parameters for both low and high-resolution mesh sizes. Seventy percent of this dataset is used for training, fifteen percent for validation, and fifteen percent for testing. Subsequently, we construct the EDSR network (explained in Section~\ref{sec:network}) to predict HR topologies from their LR counterparts. The model is trained with the created dataset, and the model's parameters are saved for later use (see Section~\ref{sec:training}).

The trained network's parameters are utilized to assess the performance of the proposed approach. To evaluate the effectiveness of the proposed strategy, we employed 150 samples from the testing set that were not part of the training phase. Comparisons are made between the computational costs of the proposed approach and those of SIMP-based TO. Accuracy is evaluated using four metrics: average volume fraction error, objective function error, pixel values error (MSE), and Intersection over Union (IoU).

The model's efficiency has been assessed by comparing the average computation time of our approach with the conventional SIMP-based approach. In Table~\ref{tab:time_BM}, we present the average computational time comparison between our proposed approach and the conventional method, revealing that the conventional method consumes substantially more time than the proposed method. Our approach operates at an impressive speed of only 24 seconds, considering the computation cost to generate LR topology and the prediction cost. Remarkably, the computational time of our proposed method is merely 5.25 \% of that required by the conventional method. Furthermore, it's noteworthy that once the neural network is well-trained, it can generate a near-optimal HR topology from the LR original without iterations.
\begin{table}
\centering
\caption{Average computation time comparison between the conventional method and our proposed approach for maximizing the bulk modulus. Our approach significantly reduces computation cost by 94\%.}
\begin{tabular}[t]{lccc}
\hline
& Conventional method & Proposed approach & Reduction (\%) \\
\hline
Average time& 457 sec & 24 sec & 94\\
\hline
\end{tabular}\label{tab:time_BM}
\end{table}%

\begin{figure}
    \centering
    \includegraphics{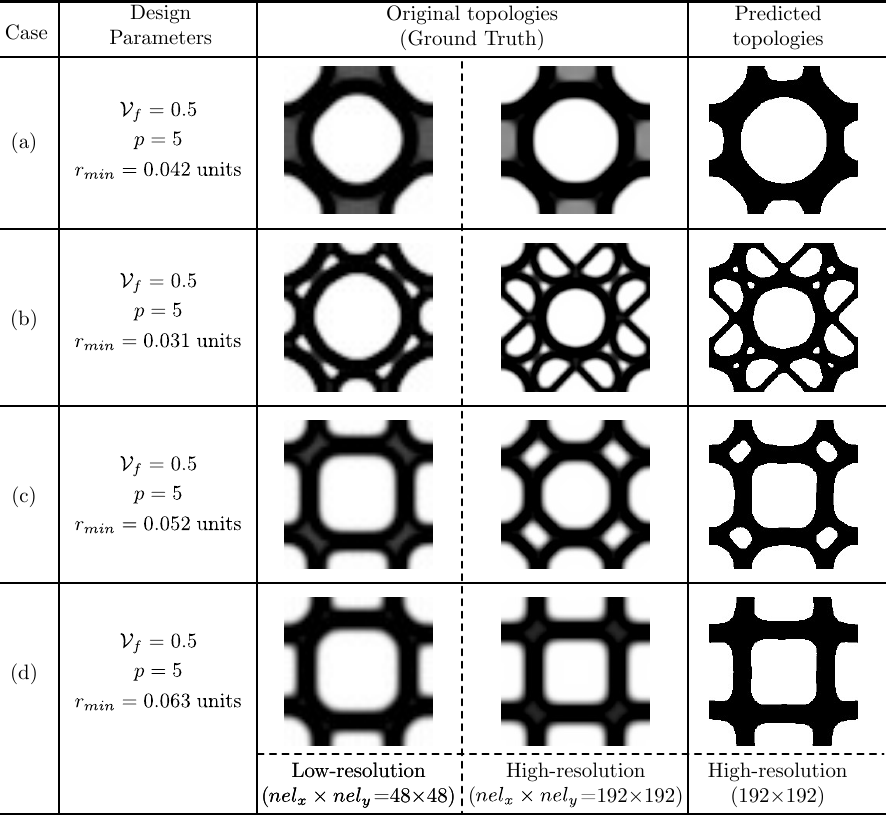}
    \caption{Comparison of optimized topologies for maximization of the bulk modulus objective function. Depicting design parameters alongside the corresponding original optimized low- and high-resolution topologies and their predicted high-resolution counterparts utilizing the proposed approach. Notably, our model accurately predicts the presence of voids, even when they are not clearly visible in the low-resolution topology, as evidenced in cases (b) and (c).}
    \label{fig:result_BM}
\end{figure}

Fig.~\ref{fig:result_BM} provides a visual representation of the comparative analysis conducted on optimized topologies to maximize the bulk modulus. Design parameters, integral to the optimization process, are systematically depicted alongside their corresponding original low- and high-resolution topologies. In addition, the figure includes the predicted high-resolution counterparts generated by our proposed approach. An essential observation from the figure is the model's remarkable accuracy in predicting the presence of voids within the metamaterial topologies. This is particularly noteworthy in cases (b) and (c), where voids are accurately predicted by the model, even when they are not distinctly visible in the LR topology. The model's capability to infer intricate details, especially in regions where they may not be readily apparent in LR structures, underscores its efficacy in enhancing the resolution of metamaterial topologies.

\begin{figure}
    \centering
    \includegraphics{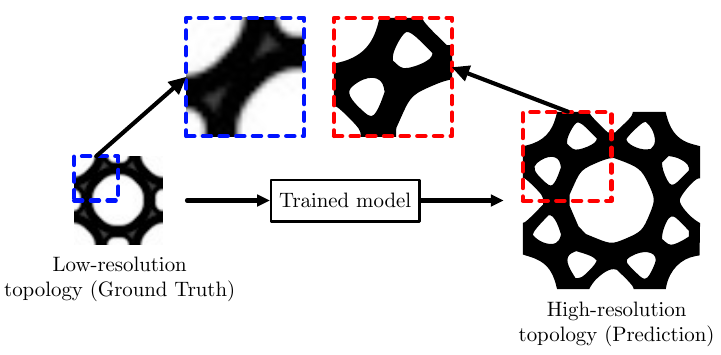}
    \caption{An illustrative comparison between the input and scaled topology is presented. Notably, the model is able to predict holes, even in instances where they are not distinctly visible in the ground truth.}
    \label{fig:holes}
\end{figure}

A comparative analysis between the input and scaled topology is demonstrated in Fig.~\ref{fig:holes}. The emphasis lies in the model's ability to discern and highlight holes, even when they may not be clearly visible in the ground truth. The visual representation provides a scientific insight into the model's proficiency in enhancing the resolution and accurately predicting structural features, contributing to understanding its capabilities in topology optimization.
The visual comparisons within the figure provide valuable insights into the model's predictive accuracy, demonstrating its capacity to capture nuanced features and generate high-resolution representations that align closely with the true structural characteristics. This visual evidence reinforces the quantitative metrics presented in the preceding theory, affirming the proficiency and reliability of the proposed approach for topology optimization in maximizing the bulk modulus objective function.

To evaluate the accuracy of our trained model, we utilized metrics such as mean squared error (MSE), intersection over union (IoU), objective function error, and volume fraction error. In Table~\ref{tab:error_BM}, we present the average prediction errors across various parameters, including pixel values (MSE), volume fraction, IoU, and objective function. The pixel errors, assessed through MSE, yielded a value of 2.24\%. This metric reflects the fidelity of our model in accurately reproducing pixel values. Apart from pixel-level evaluation, we explored broader parameters such as volume fraction, where the model exhibited an impressive average error of 1.95\%. This indicates the model's adeptness in capturing the overall distribution of material within the metamaterial structure. Similarly, the objective function error, a crucial metric governing the optimization process, showcased an average error of 3.03\%. This signifies the model's efficacy in optimizing metamaterial topologies according to the defined objective function, further affirming its reliability in the design process. The assessment extended to spatial concordance, as quantified by IoU, where the model demonstrated an average value of 0.92. This underscores the model's ability to precisely align its predictions with the ground truth, even in regions where structural intricacies demand a more detailed evaluation.

Collectively, these detailed results provide a profound understanding of the high precision achieved by our proposed method across various facets of topology generation. Importantly, this precision is achieved with a minimal computational footprint, affirming the efficiency and practicality of our approach in the realm of metamaterial topology optimization.

\begin{table}
\centering
\caption{Average prediction errors associated with the maximization of bulk modulus, considering metrics such as volume fraction, objective function, pixel value, and Intersection over Union (IoU).}
\begin{tabular}[t]{lcccc}
\hline
&Volume Fraction& Objective function & Pixel Values (MSE) & IoU\\
\hline
Average error& 1.95\% & 3.03\% & 2.24\% &0.92\\
\hline
\end{tabular}\label{tab:error_BM}
\end{table}%

\subsection{Maximizing the shear modulus}
To test the proposed approach for the objective function of maximization of the shear modulus, we generated datasets as described in Section~\ref{sec:data_gen} using varying parameters, as we did in the previous case. As in the previous case, we employed Eq.~\ref{eq:SM} to compute the objective function for maximization of the shear modulus. Similar to the previous case, we stratified the data into three portions with a ratio of 14:3:3 for training, validation, and testing, respectively.  An EDSR network (detailed in Section~\ref{sec:network}) is constructed and trained using generated data, with parameters recorded as outlined in Section~\ref{sec:training}.

After restoring the trained model's parameters, we evaluated our method using 150 samples from the testing set. This comprehensive assessment covered accuracy and computational efficiency analyses. Additionally, we directly compared the computational cost between SIMP-based TO and our proposed approach. In Table~\ref{tab:time_Shear}, we provide a clear comparison of the average computational time for the objective function of maximizing the shear modulus, demonstrating that our method operates at an impressive speed of only 26 seconds, which stands at a mere 5.27\% of the average time required by the conventional method which was 493 seconds. 
\begin{table}
\centering
\caption{Average computation time comparison between the conventional method and our proposed approach for maximizing the shear modulus. Our approach reduces computation costs by 94\%.}
\begin{tabular}[t]{lccc}
\hline
& Conventional method & Proposed approach & Reduction (\%) \\
\hline
Average time& 493 sec & 26 sec & 94 \\
\hline
\end{tabular}\label{tab:time_Shear}
\end{table}%

\begin{figure}
    \centering
    \includegraphics{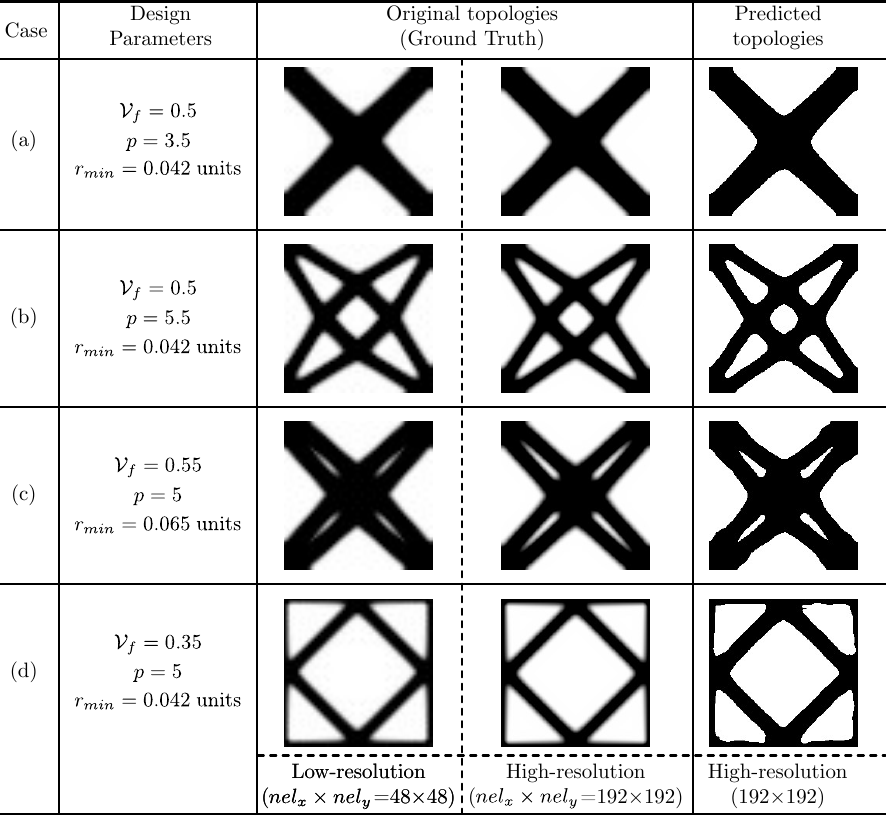}
    \caption{Comparison of optimized topologies for maximization of the shear modulus objective function. The design parameters alongside the original optimized low- and high-resolution topologies and the predicted high-resolution topologies generated using our proposed approach are shown in the figure. It can be seen that the approach is effective in smoothing the boundaries of topologies.}
    \label{fig:result_Shear}
\end{figure}

Fig.~\ref{fig:result_Shear} compares optimized designs for maximizing the shear modulus. The figure represents design parameters, original low- and high-resolution topologies, and predicted high-resolution counterparts, showcasing the model's accuracy in predicting metamaterial structures.
To assess the precision of our well-trained model, the detailed analysis presented in Table~\ref{tab:error_Shear} gives an overview of average prediction errors across diverse parameters. This comprehensive evaluation covers pixel values (MSE), volume fraction, IoU, and objective function. The obtained results underscore the exceptional accuracy of our approach, with pixel errors (MSE) registering at a mere 2.55\%. Furthermore, the average errors for the objective function, volume fraction, and IoU are remarkably low, measured at 1.57\%, 1.04\%, and 0.91, respectively. These outcomes emphasize the robust accuracy achieved by our proposed method, reinforcing its efficacy in accurately predicting and optimizing metamaterial topologies for objective function maximization of the shear modulus. Compared to the previous case of maximizing bulk modulus, the reduced complexity results in lower volume fraction error and objective function error.
\begin{table}
\centering
\caption{Average prediction errors associated with the maximization of the shear modulus, considering metrics such as volume fraction, objective function, pixel value, and Intersection over Union (IoU).}
\begin{tabular}[t]{lcccc}
\hline
&Volume Fraction&Objective function&Pixel Values&IoU\\
\hline
Average error& 1.04 \% & 1.57 \% & 2.55 \% & 0.91 \\
\hline
\end{tabular}\label{tab:error_Shear}
\end{table}%

\subsection{Maximizing the elastic modulus}
For the next study, data is generated to maximize the elastic modulus. Similar to the previous two cases, we conducted data generation and model training for maximizing the elastic modulus utilizing Eq.~\ref{eq:EM} to calculate the objective function. Using a similar process to the previous cases, topologies were generated via SIMP-based TO, resulting in a dataset of 1000 samples containing LR and HR topologies. This dataset was subsequently partitioned into training, testing, and validation subsets. An EDSR network was constructed to predict HR topologies from LR inputs. The network was trained, and the model parameters were stored. Afterward, we imported the parameters of the trained network and assessed the method's performance by utilizing 150 samples from the testing set.

As with the previous cases, we assessed both accuracy and computational efficiency. Our proposed approach consistently outperformed the conventional SIMP-based method, as demonstrated in Table~\ref{tab:time_EM}. The computational time was significantly reduced, with our method operating at just 6.39\% of the time required by the conventional approach.

\begin{table}
\centering
\caption{Comparison of average computation time between the conventional method and our proposed approach for elastic modulus maximization, showcasing a remarkable 93\% reduction in computational costs.}
\begin{tabular}[t]{lccc}
\hline
& Conventional method & Proposed approach & Reduction (\%) \\
\hline
Average time& 438 sec & 28 sec & 93 \\
\hline
\end{tabular}\label{tab:time_EM}
\end{table}%

\begin{figure}
    \centering
    \includegraphics{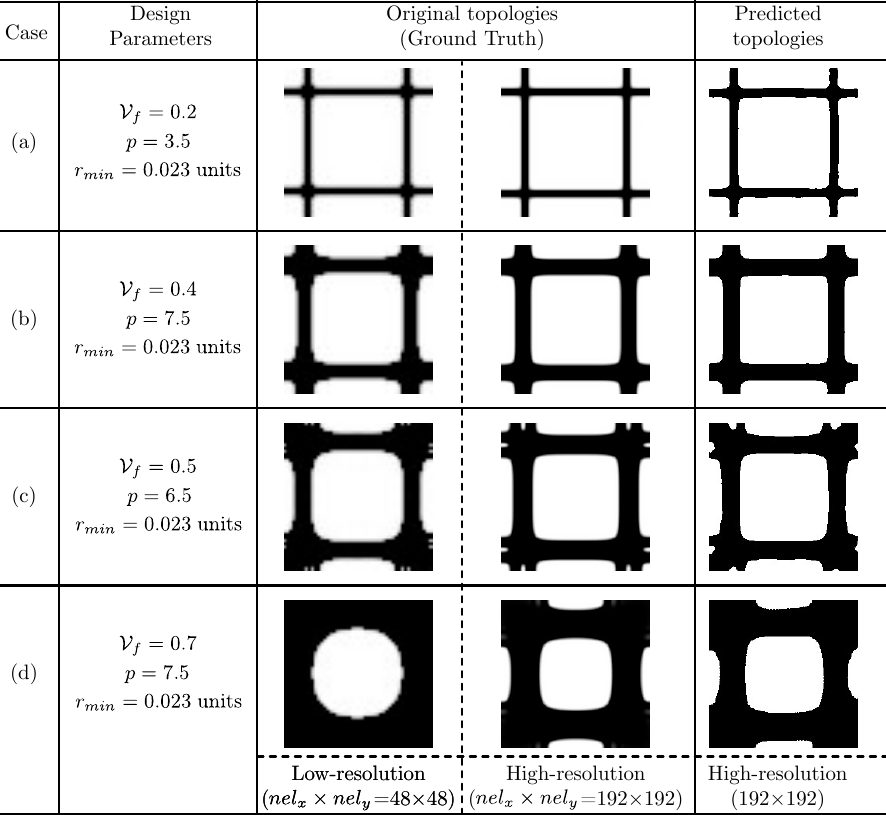}
    \caption{Illustrating the contrast between optimized designs aimed at maximizing elastic modulus. Presenting the design parameters alongside the original optimized low- and high-resolution topologies and the predicted high-resolution topologies obtained through our approach. In case (d), the input topology is unoptimized, yet our approach successfully predicts the optimized topology.}
    \label{fig:result_EM}
\end{figure}

The visual results, presented in Fig.~\ref{fig:result_EM}, confirm the accuracy of our approach, with HR topologies closely matching the optimized designs for maximizing elastic modulus objective function. A noteworthy observation pertains to case (d), where the input topology is unoptimized; nevertheless, our approach predicts the optimized topology. This highlights the model's effectiveness in identifying and improving structural features, even when the input is not initially optimized. The visual comparisons in the figure offer valuable insights into the model's predictive accuracy, confirming its capacity to generate HR topologies closely aligned with true topologies. 

To quantify the accuracy of our trained model, we employed parameters such as MSE (pixel error), IoU, objective function, and volume fraction. In Table~\ref{tab:error_BM}, we present the average prediction errors across various parameters, including pixel values (MSE), volume fraction, IoU, and objective function. The pixel errors, evaluated using MSE, yielded a value of 2.74\%. Furthermore, the average errors for the objective function, volume fraction, and IoU are calculated at 2.34\%, 1.91\%, and 0.93, respectively. These results underscore the high accuracy achieved by our proposed method in generating high-resolution topologies, all while consuming minimal computational time. The results underscore the high accuracy achieved by our method.
\begin{table}
\centering
\caption{Average prediction errors for maximization elastic modulus in context of volume fraction, objective function, pixel value, and IoU}
\begin{tabular}[t]{lcccc}
\hline
& Volume Fraction & Objective function & Pixel Values & IoU\\
\hline
Average error& 1.91 \% & 2.34 \% & 2.74 \%& 0.93\\
\hline
\end{tabular}\label{tab:error_EM}
\end{table}%

\subsection{Minimizing the Poisson's ratio}
 Our approach is further applied to design metamaterials to minimize Poisson's ratio, following the same rigorous methodology for data generation, network construction, training, and evaluation. Utilizing Eq.~\ref{eq:NPR} to calculate the objective function for this case, our method consistently surpassed the conventional approach in terms of computational efficiency, as evidenced in Table~\ref{tab:time_NPR}, with an impressive 93\% reduction in computational time. The computational cost for this objective function is recorded at 24 seconds with our approach, in contrast to the 386 seconds required by the conventional method.
\begin{table}
\centering
\caption{Computation time comparison of conventional method and our proposed approach for minimization of Poisson's ratio}
\begin{tabular}[t]{lccc}
\hline
& Conventional method & Proposed approach & Reduction (\%) \\
\hline
Average time& 386 sec & 24 sec & 93 \\
\hline
\end{tabular}\label{tab:time_NPR}
\end{table}%

\begin{figure}
    \centering
    \includegraphics{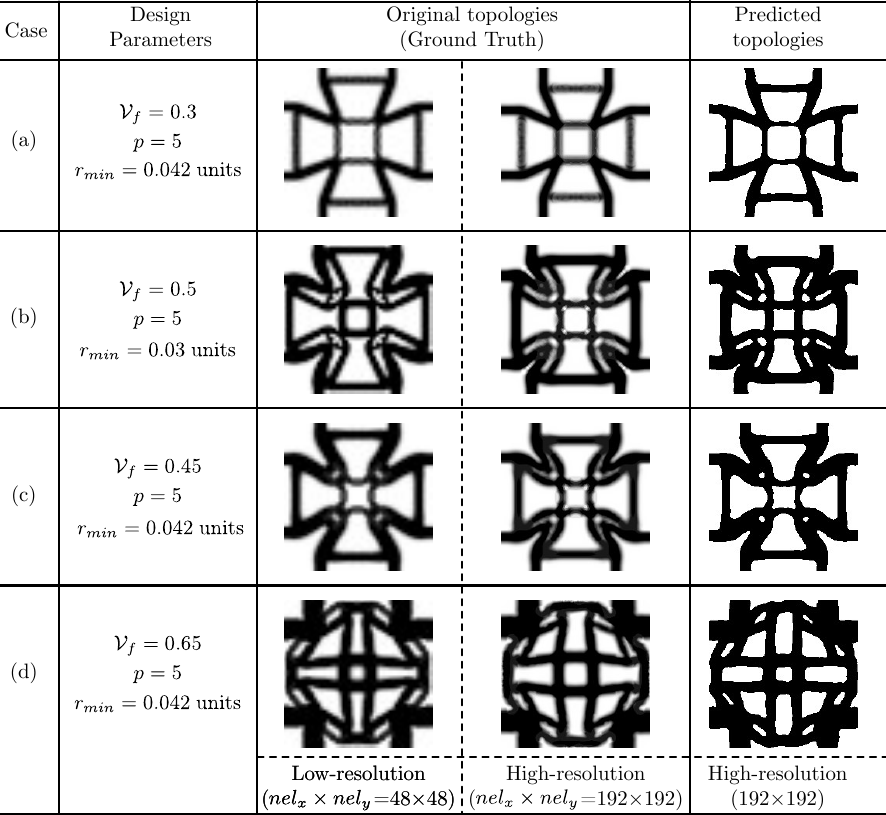}
    \caption{A comparative display of optimized designs to minimize the Poisson ratio objective function. Showcasing the design parameters alongside the original optimized low- and high-resolution topologies and the predicted high-resolution topologies through our approach. The model accurately predicts similar topologies, even when dealing with highly complex structures.}
    \label{fig:result_NPR}
\end{figure}
Visual comparisons, presented in Fig.~\ref{fig:result_NPR}, confirmed the precision of our approach in generating HR topologies designed for the objective function of minimizing the Poisson ratio.  In Table~\ref{tab:error_NPR}, one can observe the average prediction errors across various parameters, including pixel values (MSE), volume fraction, IoU, and objective function. The average errors for pixel values, volume fraction, objective function, and IoU are calculated at 6.63\%, 1.57\%, 4.24\%, and 0.84, respectively. In the case of negative Poisson ratio optimization, it's worth noting that the error rates are relatively higher compared to the previous cases due to the inherently more complex nature of the topologies involved. Achieving a negative Poisson ratio involves intricate structural arrangements that may not align as closely with conventional designs, leading to a higher degree of variation and complexity. The complexity of these topologies can challenge the predictive accuracy of any model, as achieving a negative Poisson ratio often necessitates unconventional configurations that may not fit well within the model's training data. Despite these complexities, our approach maintains its effectiveness in generating high-resolution topologies, offering a unique solution to a challenging design objective. Despite the higher error rates, these results still indicate that our method maintains a commendable level of accuracy, even in the face of intricate and unconventional designs. This underscores the robustness and adaptability of our approach in handling complex structural optimizations.

\begin{table}
\centering
\caption{Comparison of average prediction errors in the context of volume fraction, objective function, pixel value, and IoU for the objective function of minimizing the Poisson's ratio.}
\begin{tabular}[t]{lcccc}
\hline
&Volume Fraction & Objective function & Pixel Values&IoU\\
\hline
Average error& 1.57 \% & 4.24 \% & 6.63 \%& 0.84\\
\hline
\end{tabular}\label{tab:error_NPR}
\end{table}%

\section{Conclusion and future work}
\label{sec:conclusion}
In this study, a deep learning-based framework has been developed for the efficient topology optimization of mechanical metamaterials. The proposed approach combines deep learning and super-resolution techniques to reconstruct high-resolution (HR) metamaterial topologies from their low-resolution (LR) counterparts, thereby achieving substantial reductions in computational cost. A trained EDSR neural network establishes the mapping between LR and HR topologies, enabling accurate and rapid generation of optimized designs after the training phase. The training datasets are prepared using the density(SIMP)-based TO method. The major conclusions drawn from this work are as follows:
\begin{itemize}
    \item Despite the relatively limited size of the training dataset, this framework presents strong performance in prediction accuracy and computational efficiency across multiple evaluation metrics.
    \item Numerical results demonstrate significant improvements in design quality, manufacturability, and computational efficiency over conventional SIMP-based topology optimization. 
    \item The volume fraction error remained consistently below 2\%, reflecting the high precision of the model in capturing material distribution.
    \item The pixel error is contained within 3\%, reflecting the network’s ability to preserve fine geometric features.
    \item The objective function error averages around 3\%, indicating high precision in achieving the targeted mechanical properties.
    \item A high intersection over union (IoU) value (greater than 0.90) validates the network’s proficiency in reproducing spatial overlap between predicted and reference topologies.
\end{itemize}

The proposed approach also performs effectively for complex auxetic (negative Poisson’s ratio) microstructures, achieving an average volume fraction error of 1.57\%. Nevertheless, there remains scope for improvement in terms of model generalization across a wider range of metamaterial configurations. Generalizing to diverse metamaterial types is a critical concern, as overfitting the training data can lead to poor performance on new or varied metamaterial designs. Additionally, fine-tuning of hyperparameters is computationally demanding and may strongly influence performance. Although the proposed framework provides high predictive accuracy, it currently offers limited interpretability of the underlying physical mechanisms governing metamaterial behavior, which is an important aspect for engineering design. To address these limitations, future research will aim to improve model generalization and interpretability through advanced training strategies and more diverse metamaterial datasets. Integrating physics-informed neural networks (PINNs) could further mitigate overfitting and improve physical consistency between predicted and actual material responses. Extending the approach to multi-physics topology optimization could enable the design of multifunctional metamaterials with coupled mechanical, thermal, and electromagnetic responses. Overall, this study highlights the potential of deep super-resolution techniques to enhance computational efficiency, design fidelity, and manufacturability in metamaterial topology optimization.

\section*{Acknowledgment}
AS acknowledges the financial support received from the Ministry of Education, India, in the form of the Prime Minister's Research Fellows (PMRF) scholarship.

\newpage
\appendix


\begin{thebibliography}{10}

\bibitem{bendsoeGeneratingOptimalTopologies1988a}
Martin Philip Bendsøe and Noboru Kikuchi.
\newblock Generating optimal topologies in structural design using a homogenization method.
\newblock {\em Computer Methods in Applied Mechanics and Engineering}, 71(2):197–224, 1988.

\bibitem{bendsoeMaterialInterpolationSchemes1999}
Martin P. Bendsøe and Ole Sigmund.
\newblock Material interpolation schemes in topology optimization.
\newblock {\em Archive of Applied Mechanics}, 69:635–654, 1999.

\bibitem{liTopologyOptimizationConcurrent2018}
Xiaoming Li and coauthors.
\newblock Topology optimization of concurrent structures using level set methods.
\newblock {\em Computer Methods in Applied Mechanics and Engineering}, 2018.

\bibitem{ghasemiMultimaterialLevelSetbased2018}
Hamed Ghasemi et al.
\newblock Multimaterial level-set-based topology optimization.
\newblock {\em Structural and Multidisciplinary Optimization}, 2018.

\bibitem{zhangLevelSetbasedTopological2023}
Weisheng Zhang et al.
\newblock Level-set-based topological optimization in mechanics.
\newblock {\em Computer Methods in Applied Mechanics and Engineering}, 2023.

\bibitem{padhi2022efficient}
R. Padhi et al.
\newblock An efficient SIMP-based topology optimization method.
\newblock {\em Structural and Multidisciplinary Optimization}, 2022.

\bibitem{chuTopologyOptimizationMultimaterial2019}
C. Chu et al.
\newblock Topology optimization of multimaterial structures.
\newblock {\em Computer Methods in Applied Mechanics and Engineering}, 2019.

\bibitem{areiasCoupledFiniteelementTopology2021}
P. Areias et al.
\newblock Coupled finite-element and topology optimization framework.
\newblock {\em International Journal for Numerical Methods in Engineering}, 2021.

\bibitem{karuthedath2023continuous}
A. Karuthedath et al.
\newblock Continuous topology optimization for multiphysics problems.
\newblock {\em Structural and Multidisciplinary Optimization}, 2023.

\bibitem{gupta2022adaptive}
R. Gupta et al.
\newblock Adaptive SIMP approach for complex structural designs.
\newblock {\em Computer Methods in Applied Mechanics and Engineering}, 2022.

\bibitem{mattheckNewMethodStructural1990}
C. Mattheck and S. Burkhardt.
\newblock A new method for structural optimization.
\newblock {\em Engineering Optimization}, 1990.

\bibitem{baumgartnerSKOSoftKill1992}
A. Baumgartner, M. Harzheim, and M. Mattheck.
\newblock SKO: The soft kill option.
\newblock {\em Structural Optimization}, 1992.

\bibitem{huangConvergentMeshindependentSolutions2007}
X. Huang and Y. M. Xie.
\newblock Convergent and mesh-independent solutions for BESO.
\newblock {\em Engineering Computations}, 24(4):327–346, 2007.

\bibitem{huangEvolutionaryTopologyOptimization2010}
X. Huang and Y. M. Xie.
\newblock Evolutionary topology optimization of continuum structures.
\newblock {\em Structural and Multidisciplinary Optimization}, 2010.

\bibitem{shobeiri2016optimal}
S. Shobeiri et al.
\newblock Optimal design in structural optimization using heuristic approaches.
\newblock {\em Structural and Multidisciplinary Optimization}, 2016.

\bibitem{mauersberger2023topology}
K. Mauersberger et al.
\newblock Topology optimization: Modern advances and future directions.
\newblock {\em Structural and Multidisciplinary Optimization}, 2023.

\bibitem{munkBenefitsApplyingTopology2019}
David J. Munk, Douglass J. Auld, Grant P. Steven, and Gareth A. Vio.
\newblock On the benefits of applying topology optimization to structural design of aircraft components.
\newblock {\em Structural and Multidisciplinary Optimization}, 60(2):1245–1266, 2019.

\bibitem{zhuTopologyOptimizationAircraft2016a}
Ji-Hong Zhu, Wei-Hong Zhang, and Liang Xia.
\newblock Topology optimization in aircraft and aerospace structures design.
\newblock {\em Archives of Computational Methods in Engineering}, 23:595–622, 2016.

\bibitem{yoonTopologyOptimizationStationary2010}
[Author(s) Yoon …].
\newblock Topology optimization for stationary fluid-structure interaction (FSI) problems.
\newblock {\em [Journal Name]}, [volume]:[pages], 2010.

\bibitem{andreasenTopologyOptimizationFluid2013}
[Author(s) Andreasen …].
\newblock Topology optimization of fluid flows.
\newblock {\em [Journal Name]}, [volume]:[pages], 2013.

\bibitem{xiaRecentAdvancesTopology2017}
[Author(s) Xia …].
\newblock Recent advances in topology optimization of multiscale nonlinear structures.
\newblock {\em [Journal Name]}, [volume]:[pages], 2017.

\bibitem{patelImprovingConnectivityAccelerating2022}
[Author(s) Patel …].
\newblock Improving connectivity and accelerating topology optimization of multiscale nonlinear structures.
\newblock {\em [Journal Name]}, [volume]:[pages], 2022.

\bibitem{saurabh2023impact}
Saurabh [LastName], Gehani Harsh, Bhilare Sanket, Patel Sagar, and Kara Levent.
\newblock Impact of topology optimization on metamaterial microstructure designs.
\newblock {\em [Journal Name]}, [volume]:[pages], 2023.

\bibitem{liuTopologicalDesignMicrostructures2021}
Xiang Liu, Zhanli Liu, Shaoqing Cui, Chengcheng Luo, Chenfeng Li, and Zhuo Zhuang.
\newblock Topological design of microstructures using image-based modelling and deep learning.
\newblock {\em Computer Methods in Applied Mechanics and Engineering}, 347:735–753, 2019.

\bibitem{wiltAcceleratingAuxeticMetamaterial2020a}
[FirstName] Wilt and co-authors.
\newblock Accelerating auxetic metamaterial design via topology optimization.
\newblock {\em [Journal Name]}, [volume]:[pages], 2020.

\bibitem{vangelatosArchitectedMetamaterialsTailored2019b}
[FirstName] Vangelatos and co-authors.
\newblock Architected metamaterials tailored by topology optimization.
\newblock {\em [Journal Name]}, [volume]:[pages], 2019.

\bibitem{sigmundSystematicDesignMetamaterials2009}
Ole Sigmund.
\newblock Systematic design of metamaterials by topology optimization.
\newblock In {\em IUTAM Symposium on Modelling Nanomaterials and Nanosystems}, pages 151--159. Springer Netherlands, 2009. doi:10.1007/978-1-4020-9557-3\_16.

\bibitem{agrawalRobustTopologyOptimization2022}
Gourav Agrawal, Abhinav Gupta, Rajib Chowdhury, and Anupam Chakrabarti.
\newblock Robust topology optimization of negative Poisson's ratio metamaterials under material uncertainty.
\newblock {\em Finite Elements in Analysis and Design}, 198:103649, 2022. doi:10.1016/j.finel.2021.103649.

\bibitem{mizziAuxeticMetamaterialsExhibiting2015a}
Luke Mizzi, Keith M. Azzopardi, Daphne Attard, Joseph N. Grima, and Ruben Gatt.
\newblock Auxetic metamaterials exhibiting giant negative Poisson's ratios.
\newblock {\em physica status solidi (RRL) – Rapid Research Letters}, 9(7):425--430, 2015. doi:10.1002/pssr.201510178.

\bibitem{gattHierarchicalAuxeticMechanical2015b}
Ruben Gatt, Luke Mizzi, Joseph I. Azzopardi, Keith M. Azzopardi, Daphne Attard, Aaron Casha, Joseph Briffa, and Joseph N. Grima.
\newblock Hierarchical auxetic mechanical metamaterials.
\newblock {\em Scientific Reports}, 5:8395, 2015. doi:10.1038/srep08395.

\bibitem{drosopoulosMechanicalBehaviourAuxetic2016a}
Georgios A. Drosopoulos, Nikolaos Kaminakis, Nikoletta Papadogianni, and Georgios E. Stavroulakis.
\newblock Mechanical behaviour of auxetic microstructures using contact mechanics and elastoplasticity.
\newblock {\em Key Engineering Materials}, 681:100--116, 2016. doi:10.4028/www.scientific.net/KEM.681.100.

\bibitem{zhou2023ready}
Ying Zhou, Liang Gao, and Hao Li.
\newblock A ready-to-manufacture optimization design of 3D chiral auxetics for additive manufacturing.
\newblock {\em Engineering with Computers}, pages 1--22, 2023.

\bibitem{smithMetamaterialsNegativeRefractive2004}
D. R. Smith, J. B. Pendry, and M. C. K. Wiltshire.
\newblock Metamaterials and negative refractive index.
\newblock {\em Science}, 305(5685):788--792, 2004. doi:10.1126/science.1096796.

\bibitem{wangTopologicalDesignMechanical2016}
Yu Wang, Zhen Luo, Nong Zhang, and Tao Wu.
\newblock Topological design for mechanical metamaterials using a multiphase level set method.
\newblock {\em Structural and Multidisciplinary Optimization}, 54(4):937--952, 2016. doi:10.1007/s00158-016-1458-6.

\bibitem{chenHarnessingMultilayeredSoil2019}
Yanyu Chen, Feng Qian, Fabrizio Scarpa, Lei Zuo, and Xiaoying Zhuang.
\newblock Harnessing multi-layered soil to design seismic metamaterials with ultralow frequency band gaps.
\newblock {\em Materials \& Design}, 175:107813, 2019. doi:10.1016/j.matdes.2019.107813.

\bibitem{drosopoulosEvaluationDynamicResponse2019}
Georgios Drosopoulos and Preyolin Naidoo.
\newblock Evaluation of the dynamic response of structures using auxetic-type base isolation.
\newblock {\em Frattura ed Integrità Strutturale}, 14(51):52--70, 2019. doi:10.3221/IGF-ESIS.51.05.

\bibitem{chenAcousticCloakingThree2007}
Huanyang Chen and C. T. Chan.
\newblock Acoustic cloaking in three dimensions using acoustic metamaterials.
\newblock {\em Applied Physics Letters}, 91(18):183518, 2007. doi:10.1063/1.2803315.

\bibitem{noguchiLevelSetbasedTopology2022}
Yuki Noguchi, Kei Matsushima, and Takayuki Yamada.
\newblock Level set-based topology optimization for the design of labyrinthine acoustic metamaterials.
\newblock {\em Materials \& Design}, 219:110832, 2022. doi:10.1016/j.matdes.2022.110832.

\bibitem{liTopologicalDesignPentamode2021}
Zuyu Li, Zhen Luo, Lai-Chang Zhang, and Chun-Hui Wang.
\newblock Topological design of pentamode lattice metamaterials using a ground structure method.
\newblock {\em Materials \& Design}, 202:109523, 2021. doi:10.1016/j.matdes.2021.109523.


\bibitem{sigmundMaterialsPrescribedConstitutive1994}
Ole Sigmund.
\newblock Materials with prescribed constitutive parameters: An inverse homogenization problem.
\newblock {\em International Journal of Solids and Structures}, 31(17):2313--2329, 1994.
\newblock doi:10.1016/0020-7683(94)90154-6.

\bibitem{nevesOptimalDesignPeriodic2000}
M.~M. Neves, H.~Rodrigues, and J.~M. Guedes.
\newblock Optimal design of periodic linear elastic microstructures.
\newblock {\em Computers \& Structures}, 76(1–3):421--429, 2000.
\newblock doi:10.1016/S0045-7949(99)00172-8.

\bibitem{diazTopologyOptimizationMethod2010}
Alejandro~R. Diaz and Ole Sigmund.
\newblock A topology optimization method for design of negative permeability metamaterials.
\newblock {\em Structural and Multidisciplinary Optimization}, 41(2):163--177, 2010.
\newblock doi:10.1007/s00158-009-0416-y.

\bibitem{luTopologyOptimizationAcoustic2013a}
Lirong Lu, Takashi Yamamoto, Masaki Otomori, Takayuki Yamada, Kazuhiro Izui, and Shinji Nishiwaki.
\newblock Topology optimization of an acoustic metamaterial with negative bulk modulus using local resonance.
\newblock {\em Finite Elements in Analysis and Design}, 72:1--12, 2013.
\newblock doi:10.1016/j.finel.2013.04.005.

\bibitem{luoLevelSetbasedParameterization2008}
Zhen Luo, Michael~Yu Wang, Shengyin Wang, and Peng Wei.
\newblock A level set-based parameterization method for structural shape and topology optimization.
\newblock {\em International Journal for Numerical Methods in Engineering}, 76(1):1--26, 2008.
\newblock doi:10.1002/nme.2092.

\bibitem{huangTopologicalDesignMicrostructures2011a}
X. Huang, A. Radman, and Y.~M. Xie.
\newblock Topological design of microstructures of cellular materials for maximum bulk or shear modulus.
\newblock {\em Computational Materials Science}, 50(6):1861--1870, 2011.
\newblock doi:10.1016/j.commatsci.2011.01.030.

\bibitem{amir2010efficient}
Oded Amir, Mathias Stolpe, and Ole Sigmund.
\newblock Efficient use of iterative solvers in nested topology optimization.
\newblock {\em Structural and Multidisciplinary Optimization}, 42:55--72, 2010.
\newblock doi:10.1007/s00158-009-0469-z.

\bibitem{wang2007large}
Shun Wang, Eric de~Sturler, and Glaucio~H. Paulino.
\newblock Large-scale topology optimization using preconditioned {Krylov} subspace methods with recycling.
\newblock {\em International Journal for Numerical Methods in Engineering}, 69(12):2441--2468, 2007.
\newblock doi:10.1002/nme.1850.

\bibitem{borrvall2001large}
Thomas Borrvall and Joakim Petersson.
\newblock Large-scale topology optimization in 3D using parallel computing.
\newblock {\em Computer Methods in Applied Mechanics and Engineering}, 190(46–47):6201--6229, 2001.
\newblock doi:10.1016/S0045-7825(01)00226-8.

\bibitem{wang2021deep}
Chunpeng Wang, Song Yao, Zhangjun Wang, and Jie Hu.
\newblock Deep super-resolution neural network for structural topology optimization.
\newblock {\em Engineering Optimization}, 53(12):2108--2121, 2021.
\newblock doi:10.1080/0305215X.2020.1808431.

\bibitem{lecun2015deep}
Yann LeCun, Yoshua Bengio, and Geoffrey Hinton.
\newblock Deep learning.
\newblock {\em Nature}, 521(7553):436--444, 2015.
\newblock doi:10.1038/nature14539.

\bibitem{jiang2022auto}
Sheng Jiang, Zifeng Cheng, Lei Yang, and Luming Shen.
\newblock An auto-tuned hybrid deep learning approach for predicting fracture evolution.
\newblock {\em Engineering with Computers}, pages 1--18, 2022.
\newblock doi:10.1007/s00366-022-01699-2.

\bibitem{manikkan2023transfer}
Sreehari Manikkan and Balaji Srinivasan.
\newblock Transfer physics informed neural network: A new framework for distributed physics informed neural networks via parameter sharing.
\newblock {\em Engineering with Computers}, 39(4):2961--2988, 2023.
\newblock doi:10.1007/s00366-022-01867-4.

\bibitem{gantovnik2006multi}
Vladimir Gantovnik, Santosh Tiwari, Georges Fadel, and Yi Miao.
\newblock Multi-objective vehicle layout optimization.
\newblock In {\em 11th AIAA/ISSMO Multidisciplinary Analysis and Optimization Conference}, page 6978, 2006.
\newblock doi:10.2514/6.2006-6978.

\bibitem{yan2023real}
Jun Yan, Dongling Geng, Qi Xu, and Haijiang Li.
\newblock Real-time topology optimization based on convolutional neural network by using retrain skill.
\newblock {\em Engineering with Computers}, pages 1--15, 2023.
\newblock doi:10.1007/s00366-023-01873-2.

\bibitem{meng2017ultrasonic}
Min Meng, Yiting~Jacqueline Chua, Erwin Wouterson, and Chin~Peng~Kelvin Ong.
\newblock Ultrasonic signal classification and imaging system for composite materials via deep convolutional neural networks.
\newblock {\em Neurocomputing}, 257:128--135, 2017.
\newblock doi:10.1016/j.neucom.2016.11.084.

\bibitem{wang2018model}
Zheng Wang, Dunhui Xiao, Fangxin Fang, Rajesh Govindan, Christopher~C. Pain, and Yike Guo.
\newblock Model identification of reduced order fluid dynamics systems using deep learning.
\newblock {\em International Journal for Numerical Methods in Fluids}, 86(4):255--268, 2018.
\newblock doi:10.1002/fld.4482.

\bibitem{linden2023neural}
Lennart Linden, Dominik~K. Klein, Karl~A. Kalina, J{\"o}rg Brummund, Oliver Weeger, and Markus K{\"a}stner.
\newblock Neural networks meet hyperelasticity: A guide to enforcing physics.
\newblock {\em Journal of the Mechanics and Physics of Solids}, 174:105363, 2023.
\newblock doi:10.1016/j.jmps.2023.105363.

\bibitem{park2003super}
Seong~Jin Park, Min~Kyu Park, and Moon~Gi Kang.
\newblock Super-resolution image reconstruction: A technical overview.
\newblock {\em IEEE Signal Processing Magazine}, 20(3):21--36, 2003.
\newblock doi:10.1109/MSP.2003.1203207.

\bibitem{greenspan2009super}
Hayit Greenspan.
\newblock Super-resolution in medical imaging.
\newblock {\em The Computer Journal}, 52(1):43--63, 2009.
\newblock doi:10.1093/comjnl/bxm075.

\bibitem{li2015super}
Shen Li, Lihong Yuan, Jinglei Zhou, and Jianjun Chen.
\newblock Super-resolution reconstruction algorithm for remote sensing images based on sparse representation.
\newblock {\em IEEE Transactions on Geoscience and Remote Sensing}, 53(3):1440--1452, 2015.
\newblock doi:10.1109/TGRS.2014.2330427.

\bibitem{dong2014learning}
Chao Dong, Chen~Change Loy, Kaiming He, and Xiaoou Tang.
\newblock Learning a deep convolutional network for image super-resolution.
\newblock In {\em European Conference on Computer Vision (ECCV)}, pages 184--199. Springer, 2014.
\newblock doi:10.1007/978-3-319-10593-2\_13.

\bibitem{he2016deep}
Kaiming He, Xiangyu Zhang, Shaoqing Ren, and Jian Sun.
\newblock Deep residual learning for image recognition.
\newblock In {\em Proceedings of the IEEE Conference on Computer Vision and Pattern Recognition (CVPR)}, pages 770--778, 2016.
\newblock doi:10.1109/CVPR.2016.90.

\bibitem{kim2016accurate}
Jiwon Kim, Jung Kwon Lee, and Kyoung Mu Lee.
\newblock Accurate image super-resolution using very deep convolutional networks.
\newblock In {\em Proceedings of the IEEE Conference on Computer Vision and Pattern Recognition (CVPR)}, pages 1646--1654, 2016.
\newblock doi:10.1109/CVPR.2016.182.

\bibitem{ledig2017photo}
Christian Ledig, Lucas Theis, Ferenc Huszár, Jose Caballero, Andrew Cunningham, Alejandro Acosta, Andrew Aitken, Alykhan Tejani, Johannes Totz, Zehan Wang, and Wenzhe Shi.
\newblock Photo-realistic single image super-resolution using a generative adversarial network.
\newblock In {\em Proceedings of the IEEE Conference on Computer Vision and Pattern Recognition (CVPR)}, pages 4681--4690, 2017.
\newblock doi:10.1109/CVPR.2017.19.

\bibitem{lim2017enhanced}
Bee Lim, Sanghyun Son, Heewon Kim, Seungjun Nah, and Kyoung Mu Lee.
\newblock Enhanced deep residual networks for single image super-resolution.
\newblock In {\em Proceedings of the IEEE Conference on Computer Vision and Pattern Recognition Workshops (CVPRW)}, pages 136--144, 2017.
\newblock doi:10.1109/CVPRW.2017.151.

\bibitem{hassaniReviewHomogenizationTopology1998}
Behrooz Hassani and Esmaeel Hinton.
\newblock A review of homogenization and topology optimization — I. Homogenization theory for media with periodic structure.
\newblock {\em Computers \& Structures}, 69(6):707--717, 1998.
\newblock doi:10.1016/S0045-7949(98)00131-6.

\bibitem{xiaDesignMaterialsUsing2015}
Liang Xia and Piotr Breitkopf.
\newblock Design of materials using topology optimization and energy-based homogenization approach in MATLAB.
\newblock {\em Structural and Multidisciplinary Optimization}, 52(6):1229--1241, 2015.
\newblock doi:10.1007/s00158-015-1294-0.

\bibitem{zhangUsingStrainEnergybased2007}
Weihong Zhang, Weizhong Guo, and Xu Guo.
\newblock Using strain energy-based homogenization method to design microstructures of composites for desired properties.
\newblock {\em Computational Materials Science}, 39(1):159--165, 2007.
\newblock doi:10.1016/j.commatsci.2006.05.008.

\bibitem{sigmundMorphologybasedBlackWhite2007}
Ole Sigmund.
\newblock Morphology-based black and white filters for topology optimization.
\newblock {\em Structural and Multidisciplinary Optimization}, 33(4–5):401--424, 2007.
\newblock doi:10.1007/s00158-006-0087-x.


\bibitem{sigmundNumericalInstabilitiesTopology1998}
Ole Sigmund and Joakim Petersson.
\newblock Numerical instabilities in topology optimization: A survey on procedures dealing with checkerboards, mesh-dependencies, and local minima.
\newblock {\em Structural Optimization}, 16(1):68--75, 1998.
\newblock doi:10.1007/BF01214002.

\bibitem{andreassenEfficientTopologyOptimization2011}
Erik Andreassen, Anders Clausen, Mattias Schevenels, Boyan S. Lazarov, and Ole Sigmund.
\newblock Efficient topology optimization in MATLAB using 88 lines of code.
\newblock {\em Structural and Multidisciplinary Optimization}, 43(1):1--16, 2011.
\newblock doi:10.1007/s00158-010-0594-7.

\bibitem{bendsoeOptimizationStructuralTopology1995}
Martin P. Bendsoe.
\newblock Optimization of Structural Topology, Shape, and Material.
\newblock {\em Springer-Verlag}, Berlin Heidelberg, 1995.
\newblock doi:10.1007/978-3-662-03115-5.


\bibitem{gu2018recent}
Jiuxiang Gu, Zhenhua Wang, Jason Kuen, Lianyang Ma, Amir Shahroudy, Bing Shuai, Ting Liu, Xingxing Wang, Gang Wang, Jianfei Cai, and Tsuhan Chen.
\newblock Recent advances in convolutional neural networks.
\newblock {\em Pattern Recognition}, 77:354--377, 2018.
\newblock doi:10.1016/j.patcog.2017.10.013.

\bibitem{developers2022tensorflow}
TensorFlow Developers.
\newblock TensorFlow: Large-scale machine learning on heterogeneous systems.
\newblock Software available from \url{https://www.tensorflow.org}, 2022.

\bibitem{nair2010rectified}
Vinod Nair and Geoffrey E. Hinton.
\newblock Rectified linear units improve restricted boltzmann machines.
\newblock In {\em Proceedings of the 27th International Conference on Machine Learning (ICML-10)}, pages 807--814, 2010.

\bibitem{he2015delving}
Kaiming He, Xiangyu Zhang, Shaoqing Ren, and Jian Sun.
\newblock Delving deep into rectifiers: Surpassing human-level performance on ImageNet classification.
\newblock In {\em Proceedings of the IEEE International Conference on Computer Vision (ICCV)}, pages 1026--1034, 2015.

\bibitem{mitchell1997machine}
Tom M. Mitchell.
\newblock Machine Learning.
\newblock {\em McGraw-Hill}, New York, 1997.
\newblock ISBN: 978-0070428072.

\bibitem{shi2016real}
Wenzhe Shi, Jose Caballero, Ferenc Huszár, Johannes Totz, Andrew P. Aitken, Rob Bishop, Daniel Rueckert, and Zehan Wang.
\newblock Real-time single image and video super-resolution using an efficient sub-pixel convolutional neural network.
\newblock In {\em Proceedings of the IEEE Conference on Computer Vision and Pattern Recognition (CVPR)}, pages 1874--1883, 2016.
\newblock doi:10.1109/CVPR.2016.207.

\bibitem{kingma2014adam}
Diederik P. Kingma and Jimmy Ba.
\newblock Adam: A method for stochastic optimization.
\newblock {\em arXiv preprint arXiv:1412.6980}, 2014.
\newblock \url{https://arxiv.org/abs/1412.6980}.

\bibitem{sun2020optimization}
Ruo-Yu Sun.
\newblock Optimization for deep learning: An overview.
\newblock {\em Journal of the Operations Research Society of China}, 8(2):249--294, 2020.
\newblock doi:10.1007/s40305-020-00309-6.

\bibitem{clark2015pillow}
Alex Clark.
\newblock Pillow (PIL Fork) Documentation.
\newblock {\em Read the Docs}, 2015.
\newblock Available at: \url{https://buildmedia.readthedocs.org/media/pdf/pillow/latest/pillow.pdf}.

\bibitem{gupta55lineCodeLargescale2020}
Abhinav Gupta, Rajib Chowdhury, Anupam Chakrabarti, and Timon Rabczuk.
\newblock A 55-Line Code for Large-Scale Parallel Topology Optimization in 2D and 3D.
\newblock {\em arXiv preprint arXiv:2012.08208}, 2020.
\newblock \url{https://arxiv.org/abs/2012.08208}.
\newblock doi:10.48550/arXiv.2012.08208.

\bibitem{AutomatedSolutionDifferential}
Anders Logg, Kent-Andre Mardal, and Garth N. Wells (editors).
\newblock {\em Automated Solution of Differential Equations by the Finite Element Method: The FEniCS Book}.
\newblock Springer, Berlin, Heidelberg, 2012.
\newblock doi:10.1007/978-3-642-23099-8.

\bibitem{alnaesFEniCSProjectVersion2015}
Martin Aln{\ae}s, Jan Blechta, Johan Hake, August Johansson, Benjamin Kehlet, Anders Logg, Chris Richardson, Johannes Ring, Marie E. Rognes, and Garth N. Wells.
\newblock The FEniCS Project Version 1.5.
\newblock {\em Archive of Numerical Software}, 3(100), 2015.
\newblock doi:10.11588/ans.2015.100.20553.

\end{thebibliography}
\end{document}